\documentclass[preprint,11pt,number,sort&compress]{elsarticle}
\usepackage[T1]{fontenc}
\usepackage[margin=0.96in,top=0.89in,bottom=0.91in,letterpaper]{geometry}
\usepackage{amsmath,amssymb,amsthm,mathtools,newtxtext,newtxmath}
\usepackage{bm,enumitem}
\newcounter{qmalg}
\usepackage{microtype,graphicx,booktabs,xcolor}
\usepackage[font=small,labelfont=bf,labelsep=period]{caption}
\usepackage{needspace,flafter,placeins}

\definecolor{linkblue}{HTML}{244963}
\usepackage[colorlinks=true,linkcolor=linkblue,citecolor=linkblue,urlcolor=linkblue]{hyperref}
\hypersetup{pdftitle={Quantum multiscale algorithms for the nonlinear dynamics of material defects},pdfauthor={Xiantao Li}}
\newtheorem{proposition}{Proposition}[section]
\newtheorem{theorem}[proposition]{Theorem}
\theoremstyle{definition}

\theoremstyle{plain}
\journal{Computer Methods in Applied Mechanics and Engineering}
\makeatletter
\def\ps@pprintTitle{\let\@oddhead\@empty\let\@evenhead\@empty
 \def\@oddfoot{\footnotesize\itshape Preprint\hfill\normalfont\thepage}%
 \let\@evenfoot\@oddfoot}
\makeatother
\begin{document}
\begin{frontmatter}
\title{Quantum multiscale algorithms for the nonlinear dynamics of material defects}
\author[psu]{Xiantao Li\corref{cor}}
\ead{xxl12@psu.edu}
\cortext[cor]{Corresponding author.}
\affiliation[psu]{organization={Department of Mathematics, The Pennsylvania State University},
  city={University Park}, state={PA},
  postcode={16802}, country={USA}}

\begin{abstract}
Simulating a crystal defect requires advancing both a local nonlinear core and many surrounding atoms because the two regions interact repeatedly. We present two quantum multiscale algorithms that couple a nonlinear atomistic core to a harmonic exterior while retaining the lattice dispersion. In concurrent coupling, classical molecular dynamics for the atoms at the defects supplies the interface displacement for the surrounding atoms, while a quantum simulation of the driven exterior returns the traction. In the offline--online algorithm, the quantum calculation measures the exterior response to prescribed interface impulses in advance. The classical core then uses this response for each new displacement.

In the quantum computation of the interface response,  we use shared time averages and recover all $N_T$ normalized responses to accuracy $\eta$, only using $\widetilde O\big((1+\alpha T)/\eta\big)$ quantum queries. Here $T$ is the observation time, $\alpha$ bounds the highest exterior frequency, and $\widetilde O$ suppresses logarithmic factors.  At fixed core and interface sizes, the quantum algorithm requires $\widetilde O((1+\alpha T)\log(2N_{\mathcal E})/\eta)$ quantum gates. The subsequent classical time-marching costs $\widetilde O(N_T)$ operations. As a comparison, the full-scale classical simulation requires $\Theta(N_TN_{\mathcal E})$ classical operations for $N_{\mathcal E}$ exterior coordinates. Numerical benchmarks quantify the readout and preparation costs. 
\end{abstract}

\end{frontmatter}

\section{Introduction}
\label{sec:intro}

Over the past few decades, atomistic/continuum coupling methods have made it possible to study individual defects in crystals subjected to loading on much larger scales. Atomic resolution is needed where a dislocation glides or a crack advances, while continuum mechanics describes how the surrounding solid deforms and transmits the applied load. For example, an early quasicontinuum calculation by Tadmor, Ortiz and Phillips reproduced the dissociation of an edge dislocation in aluminum into two Shockley partials~\cite{tadmor1996}. Subsequent work coupled molecular dynamics near a crack tip to continuum elastodynamics farther away, allowing the crack to respond to loading imposed at the specimen boundary~\cite{liyange2010}. In a three-dimensional bridging-scale calculation, Park et al. captured crack branching in a face-centered cubic lattice and compared the result with full molecular dynamics~\cite{ParkKarpovKleinLiu2005}.

In dynamics, the surrounding crystal also carries waves emitted by the defect. A lattice wave can have a small amplitude by the time it reaches the atomistic/continuum interface and still have a wavelength of only a few lattice spacings. At these wavelengths, a coarse continuum model generally has a different dispersion relation from that of the crystal~\cite{SelesonEtAl2009}. Part of the outgoing wave can therefore reflect at the computational interface and return to the defect, altering the force on the atoms whose motion is being studied. Such an artificial reflection can obscure the wave physically scattered by the defect. This difficulty motivated matching conditions that suppress phonon reflections~\cite{ehuang2001,ehuang2002}. But in general, these problems remain largely open.

The recent emergence of quantum algorithms for harmonic lattice dynamics offers a new, promising direction to treat the exterior explicitly. The amplitudes of a state with \(O(\log N_{\mathcal E})\) address qubits can encode the \(N_{\mathcal E}\) scalar displacement coordinates of the surrounding atoms~\cite{babbush2023,li2025}. The difficulty then  is reduced to the coupling of the field to a defect whose forces depend nonlinearly on its evolving bonds. In a perfect harmonic crystal a traveling Bloch mode propagates without scattering into other modes. A defect can redirect part of a packet toward its source, and the resulting interface motion radiates waves that influence the subsequent core motion. 

We consider two ways to obtain these forces. In \emph{concurrent coupling}, the defect supplies its displacement at each step and the quantum calculation measures the force exerted by the resulting wave field by signed overlap estimation.  Each new measurement can also require replaying the interface motion that prepared it. The two regions stay coupled during the simultaneous time integration. In contrast,  the second algorithm uses \emph{offline--online dynamic substructuring}. Although the future displacement is unknown, the response to a unit interface impulse is fixed by the exterior crystal, the cut and the outer boundary. Therefore,  we can measure this response offline once and for all. During the online classical time marching of the defects, each new displacement enters the force history through the precomputed response property. 

Wagner and Liu used a projection onto continuum shape functions to separate coarse motion from atomic fluctuations in their bridging-scale method~\cite{wagner2003}. Other atomistic--continuum schemes have controlled coupling and coarsening errors near a defect and enforced kinematic compatibility in an overlap~\cite{luskin2013,xiao2004bridging}. For a harmonic crystal, Wagner, Karpov and Liu computed time-history boundary kernels for planar cuts~\cite{wagner2004planar}. Lattice Green functions, nonreflecting boundaries and rational approximations of dynamic Dirichlet-to-Neumann (DtN) maps offer other ways to obtain the returning force~\cite{cai2000,karpov2005,park2005nonreflecting,wu2018}. We derive the DtN map both in continuous time and for the same Verlet step used in the core. The coupling requires no separation of time scales and admits prescribed initial displacement and velocity, including a mechanical pulse or a realization of a classical harmonic reservoir~\cite{kantorovich2008}.

On the quantum algorithm side, Babbush et al. showed how to estimate a kinetic-energy fraction for a network of harmonic oscillators and obtained an oracle separation for a family of networks and that observable~\cite{babbush2023}. A factorization of matrix-valued force constants extends the Hamiltonian construction to crystalline lattices beyond nearest neighbors and scalar displacements~\cite{li2025}. Related work treats scalar and elastic waves, prescribed sources and interfaces, and includes a processor demonstration of one-dimensional wave dynamics~\cite{costa2019,schade2024,bosch2025,jin2024physical,xu2026}. Quantum methods in computational mechanics have also been studied for representative-volume problems and repeated microscopic finite-element solves~\cite{liu2024quantummechanics,wang2026qafe}. Quantum treatments of nonlinear evolution include nonlinear Schr\"odingerization~\cite{muraleedharan2025} and Carleman linearization in a dissipative regime~\cite{liu2021dissipative}. Here the  classical MD is conservative and evaluates the nonlinear core force in full at each step, allowing large local bond distortions.

The quantum primitives follow from the force the core requests. For on-the-fly force updates, quantum signal processing implements the Verlet propagators, a linear combination of unitaries assembles the waves from the accepted interface displacements, and quantum mean estimation reads their signed projection on a force probe. For the offline-online setting where a reusable response is precomputed, Hamiltonian simulation propagates a prescribed probe, and the measured quantity is its time-averaged overlap with a second probe. We can sample this function at Chebyshev points in time and interpolate it at all the steps of the core, or share interval averages among many lags. Chebyshev moments of local spectral measures, quantum estimators for response functions and polynomial approximations of classical harmonic memory are established techniques~\cite{weisse2006,rall2020,danz2025,ZhuVenturi2018}. We use the impulse response to relate the accuracy of the quantum measurements to the displacement error in the nonlinear trajectory. When several defects share the same crystal, the response also gives the effective, memory-dependent interactions among the defect cores.

The potential quantum speedup concerns the number \(N_{\mathcal E}\) of exterior displacement coordinates. Full-scale Verlet performs \(\Theta(N_T N_{\mathcal E})\) classical work for a fixed core over \(N_T\) steps. In contrast, for \(N_{\rm pair}\) distinct load--readout pairs at the interface, our shared interval average approach only needs \(\widetilde O(N_{\rm pair}(1+\alpha T)/\eta)\) calls to supplied quantum circuits for all lags, with encoding frequency \(\alpha\) and normalized lag tolerance \(\eta\). The core subsequently advances classically in nearly linear work in \(N_T\). At fixed \(T\), bounded physical gains and \((\Delta t)^2\asymp\varepsilon\), the first trajectory costs \(\widetilde O(\varepsilon^{-1}\log(2N_{\mathcal E}))\) ideal quantum gates under logarithmic-cost coherent access and \(\widetilde O(N_T)\) classical operations for reconstruction and marching. When the initial field is locally accessible, clamped Verlet can also exploit finite-time propagation~\cite{sakamoto2026}. At fixed time this gives the same core observable in \(\widetilde O(\varepsilon^{-1/2})\) classical operations under the assumptions of Appendix~\ref{app:incoming-box}.

This paper is organized as follows. Section~\ref{sec:model} specifies the displacement and traction exchanged at the cut. Section~\ref{sec:algorithm} follows one concurrent exchange from the classical drift to the quantum force measurement and the final kick. Section~\ref{sec:precompute} derives the generalized moments from the Schr\"odinger propagator and gives quantum algorithms for estimating them before the defect moves. Section~\ref{sec:coupling} relates their cost to core-displacement accuracy. Section~\ref{sec:numerics} tests the precomputed response and concurrent force readout, and Section~\ref{sec:conclusion} discusses implementation and open questions.

\section{The defect, its harmonic surroundings and the interface}\label{sec:model}
\subsection{From full molecular dynamics to the core and exterior}\label{sec:partition}
\begin{figure}[!htb]\centering
\includegraphics[width=.96\linewidth]{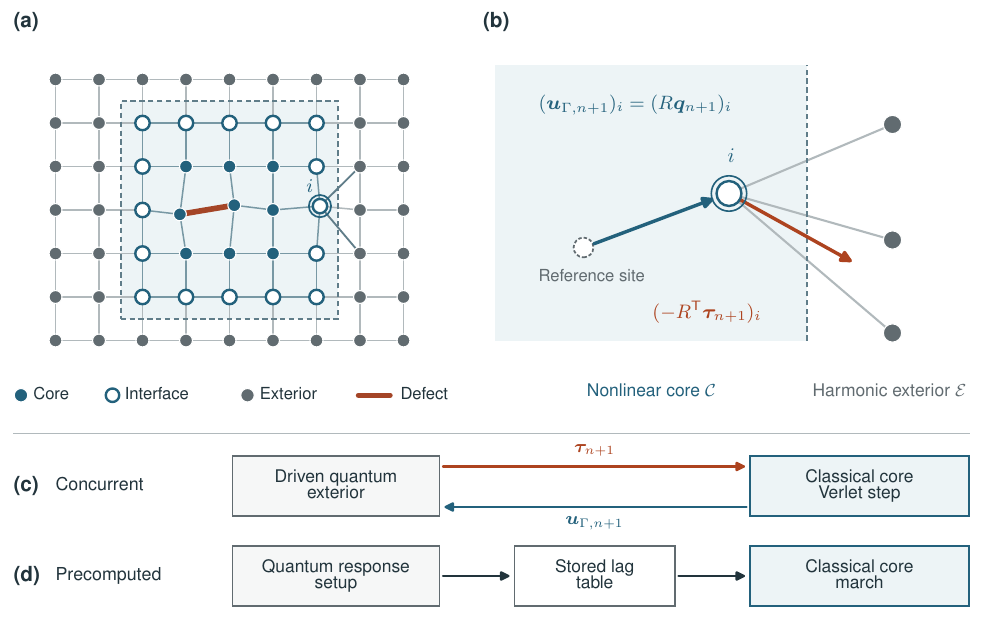}
\caption{The same crystal partition supports both coupling algorithms. (a) The dashed cut encloses the nonlinear core. Rings mark the interface atoms, and the highlighted bond contains the defect. (b) The blue arrow shows the displacement of atom \(i\) from its reference site. The orange arrow shows the cut force \(-\bm\tau_i\) on the core atom (unit masses, schematic lengths). (c) In concurrent coupling the core sends its displacement after each drift, and the measured signed traction enters the second half kick. (d) In response precomputation the quantum calculation supplies the lag table before the defect moves. The classical core uses the table at every step, together with the incident force for the chosen launch.}\label{fig:partition}
\end{figure}

We start with fully nonlinear molecular dynamics (MD). Atom \(i\), of mass \(m_i>0\), has reference position \(\bm{R}_i\), position \(\bm{r}_i\), and displacement \(\bm{d}_i=\bm{r}_i-\bm{R}_i\). For an interatomic potential \(\mathcal U\), Newton's equations for the entire crystal read
\begin{equation}
 m_i\ddot{\bm{r}}_i=-\nabla_{\bm{r}_i}\mathcal U(\bm{r}),\qquad i=1,\ldots,N .
 \label{eq:full-md}
\end{equation}
For crystalline solids, it is often not necessary to simulate the full MD model. Near a defect, a bond can stretch or rotate enough to sample the nonlinear part of \(\mathcal U\). Farther away, an elastic displacement with small gradient changes little over an interaction distance. This motivates us to retain the full potential for interactions that undergo large distortions and linearize the rest about the reference crystal, preserving its bulk phonon dispersion. Gallego and Ortiz used a related harmonic/anharmonic partition in lattice statics and represented the exterior through a lattice Green function~\cite{GallegoOrtiz1993}. In the present dynamics, we use the resolved-variable formulation~\cite{li2010coarse}, and the partition of these atoms is illustrated in Figure~\ref{fig:partition}.

The cut shown in Figure~\ref{fig:partition} partitions the atom labels into the nonlinear core \(\mathcal C\) and harmonic exterior \(\mathcal E\). The interface layer \(\Gamma\subseteq\mathcal C\) consists of the core atoms with interactions across the cut. Its thickness depends on the interaction radius. The core can have holes or consist of disjoint defect regions. We include every interface component in \(\Gamma\) and retain one common harmonic exterior \(\mathcal E\). We order the displacements with the core first. With diagonal mass matrix \(\mathsf M\) and physical momentum \(\bm p_{\rm at}=\mathsf M\dot{\bm r}\), we work with the mass-weighted displacement and momentum
\begin{equation}
 \bm u=\mathsf M^{1/2}\bm d=\binom{\bm q}{\bm y},\qquad
 \dot{\bm u}=\mathsf M^{-1/2}\bm p_{\rm at} .
 \label{eq:mass-coordinates}
\end{equation}
The crystal contains \(N=|\mathcal C|+|\mathcal E|\) atoms. With \(d\) displacement components per atom, we count \(N_{\mathcal C}=d|\mathcal C|\) core coordinates, \(N_{\mathcal E}=d|\mathcal E|\) exterior coordinates and \(N_I=d|\Gamma|\) interface coordinates. Thus \(\bm q\in\mathbb R^{N_{\mathcal C}}\), \(\bm y\in\mathbb R^{N_{\mathcal E}}\) and the prescribed trace is \(\bm u_\Gamma=R\bm q\in\mathbb R^{N_I}\). For unit masses these are the atomic displacements themselves. For several species the kinetic energy is still \(\|\dot{\bm u}\|^2/2\), with the potential evaluated at \(\bm R+\mathsf M^{-1/2}\bm u\).

\Needspace{12\baselineskip}
We expand the potential energy of the exterior and crossing interactions about a force-balanced reference. The linear term in \(\bm{y}\) vanishes, and any linear term on the interface is absorbed into the core energy \(W\). The combined potential and equations are
\begin{equation}
 V(\bm q,\bm y)=W(\bm q)+V_{\mathcal E}(\bm u_\Gamma,\bm y),\quad
 V_{\mathcal E}=\tfrac12\bm u_\Gamma^\top D\bm u_\Gamma+\bm y^\top C\bm u_\Gamma
         +\tfrac12\bm y^\top A\bm y .
 \label{eq:energy}
\end{equation}
\begin{equation}
\left\{\begin{aligned}
 \ddot{\bm q}&=-\nabla W(\bm q)-R^\top(D\bm u_\Gamma+C^\top\bm y),\\
 \ddot{\bm y}&=-A\bm y-C\bm u_\Gamma,\qquad
 \bm\tau=D_\times\bm u_\Gamma+C^\top\bm y .
\end{aligned}\right.
\label{eq:motion}
\end{equation}
The force constants give a mechanical interpretation of these blocks and fix the sign of the traction.
\begin{itemize}[leftmargin=1.45em,itemsep=2pt,topsep=3pt]
\item \(W(\bm q)\) retains the nonlinear core interactions. The restriction \(R\in\mathbb R^{N_I\times N_{\mathcal C}}\) selects the interface coordinates. The index \(a\) labels a scalar interface coordinate.
\item The physical Hessian \(\mathsf K=(K_{ij})\) has blocks \(K_{ij}=\nabla_{\bm r_i}\nabla_{\bm r_j}^{\top}\mathcal U(\bm R)\), with \(K_{ji}=K_{ij}^{\top}\). Individual off-diagonal blocks can be nonsymmetric. Translational invariance of the unrestrained crystal gives \(\sum_j K_{ij}=0\), so the linearized physical force at atom \(i\) is \(-\sum_j K_{ij}\bm d_j=\sum_{j\ne i} K_{ij}(\bm d_i-\bm d_j)\).
\item We assume harmonic stability of the clamped exterior. Its stiffness \(A=A^\top\succeq0\) contains the exterior rows and columns of the mass-weighted Hessian, including the diagonal contributions of crossing interactions. Its size is \(N_{\mathcal E}\times N_{\mathcal E}\), and its highest angular frequency is \(\Omega=\|\sqrt A\|_2=\sqrt{\lambda_{\max}(A)}\). The energy factor \(B\) in Section~\ref{sec:portpower} and its Hermitian generator \(H\) in Section~\ref{sec:ct} have the same norm, \(\|B\|_2=\|H\|_2=\Omega\).
\item The exterior rows \(j\in\mathcal E\) and interface columns \(i\in\Gamma\) form \(C\in\mathbb R^{N_{\mathcal E}\times N_I}\), with blocks \(C_{ji}=K_{ji}/\sqrt{m_jm_i}\). Its columns give the load patterns in \(-C\bm u_\Gamma\). For response precomputation we write \(N_{\rm pair}\) for the number of distinct pairs of loading and readout probes in the time-dependent lag tables. A dense reciprocal interface response has \(N_I(N_I+1)/2\) independent entries. Prescribed incident fields can require additional pairs (Section~\ref{sec:ct}).
\item The interface stiffness \(D=\nabla^2_{\bm u_\Gamma}V_{\mathcal E}\in\mathbb R^{N_I\times N_I}\) contains the Hessian of the assigned exterior and crossing interactions on their interface coordinates. In terms of full force constants, \(D_{ii'}=K_{ii'}/\sqrt{m_i m_{i'}}-[\nabla^2W(\bm 0)]_{ii'}\) for \(i,i'\in\Gamma\). To obtain the force transmitted across the cut, we keep only \(j\in\mathcal E\) in the relative-displacement force sum above. We define \(\bm\tau_i\) as the nodal holding traction in mass-weighted coordinates. The cut force on core atom \(i\) is \(-\sqrt{m_i}\,\bm\tau_i\), so
\begin{equation}
 \sqrt{m_i}\,\bm\tau_i=\sum_{j\in\mathcal E} K_{ij}(\bm d_j-\bm d_i).
 \label{eq:cut-force}
\end{equation}
Here \(C^\top\bm y\) supplies the exterior ends of the cut interactions. Their core ends give the block-diagonal matrix \(D_\times\) with \((D_\times)_{ii}=-m_i^{-1}\sum_{j\in\mathcal E}K_{ij}\). A common translation gives zero traction. For crossing pair springs with all core--core springs assigned to \(W\), \(D=D_\times\). A many-body interaction crossing the cut can couple two distinct interface atoms in \(D\), whereas \(D_\times\) is block diagonal. The core adds the known correction \((D-D_\times)\bm u_\Gamma\) to the holding traction to recover the full interface gradient of \(V_{\mathcal E}\).
\end{itemize}

Due to the underlying periodic lattice structure, each bulk force-constant block depends only on the two basis atoms and their cell offset~\cite{ashcroft1976}. One precomputed table of small matrices thus supplies all translated copies in \(A\) and the unmodified entries of \(C\). This is particularly useful for a quantum circuit, which can select an offset, shift a cell address of \(O(\log(2N_{\mathcal E}))\) bits and test membership in the exterior. The same coefficients enter the strain readout in Section~\ref{sec:portpower} and the quantum generator in Section~\ref{sec:ct}.

\paragraph{Time integration}
Our reference calculation advances every atom of \eqref{eq:motion} by velocity Verlet. For \(\bm{v}=\dot{\bm{u}}\) and mass-weighted acceleration \(\bm{a}=-\nabla V\), the updates are
\begin{equation}
 \bm{v}_{n+1/2}=\bm{v}_n+\tfrac{\Delta t}{2} \bm{a}(\bm{u}_n),\qquad
 \bm{u}_{n+1}=\bm{u}_n+\Delta t\,\bm{v}_{n+1/2},\qquad
 \bm{v}_{n+1}=\bm{v}_{n+1/2}+\tfrac{\Delta t}{2}\bm{a}(\bm{u}_{n+1}).
 \label{eq:full-verlet}
\end{equation}
For the harmonic modes of the linearization, velocity Verlet is stable when \(\Delta t\,\omega_{\max}<2\). At a stable equilibrium, \(\omega_{\max}^2=\lambda_{\max}(\nabla^2 V)\) is the Lipschitz constant of the linearized mass-weighted acceleration. The exterior alone requires \(\Delta t\,\Omega<2\). The Debye angular frequency and its square give bulk scales for acoustic frequencies and stiffness~\cite{debye1912}, although optical branches or a stiff defect can raise \(\omega_{\max}\). We compare the methods at \(t_n=n\Delta t\), \(0\le n\le N_T=T/\Delta t\), with the same initial data and outer boundary.

\paragraph{Accuracy of the local harmonic approximation}
For a translation-invariant interatomic potential, translating an entire force stencil leaves its atomic forces unchanged. The displacements that control a local linearization are therefore the changes of relative atomic positions over that stencil~\cite[Ch.~22]{ashcroft1976}. For an equilibrated, strongly stable point defect in a single-species Bravais crystal, Ehrlacher, Ortner and Shapeev obtain an \(O(R_{\mathcal C}^{-3d/2})\) error in the \(\ell^2\) norm of displacement differences when the nonlinear core contains a sufficiently large ball of radius \(R_{\mathcal C}\) in dimensions \(d=2,3\)~\cite[Theorem~4]{ehrlacher2016}. Higher-order far-field expansions give improved boundary data for such equilibrium problems~\cite{braun2025farfield}. Our coupling estimates compare solvers on the same harmonic-exterior model.

\subsection{The displacement and traction exchanged at the material cut}\label{sec:portpower}
Suppose the core has supplied the positions of its interface atoms up to time \(t\). The harmonic exterior then has a prescribed load, \(-C\bm u_\Gamma\), in \eqref{eq:motion}. Its initial displacement and velocity and this load determine \(\bm y(t)\). To advance the defect the core needs only \(C^\top\bm y(t)\). It adds \(D_\times\bm u_\Gamma(t)\) to obtain the holding traction across the cut. This displacement input and traction output define the coupling. In the next section, we show that Algorithm~\ref{alg:march} exchanges these quantities during the trajectory, while in Algorithm~\ref{alg:reusable} the core evaluates the same displacement-to-traction law from a pre-computed response (Fig.~\ref{fig:partition}c,d).

The mechanical interpretation of the readout comes from factoring the elastic energy. For the scalar wave equation, Costa, Jordan and Ostrander used a weighted incidence matrix that maps site values to edge differences~\cite{costa2019}. For oscillator networks, Babbush et al. used spring extensions weighted by the square roots of their stiffnesses~\cite{babbush2023}. For crystal interactions extending over several neighbor shells, the analogous construction factors the matrix-valued Fourier symbol of the bulk stiffness by a Fej\'er--Riesz or sum-of-squares representation~\cite{li2025}. Once available, its finite stencil supplies translated copies of the same energy coordinates throughout the lattice.

We will work with a factorization of this form in the exterior and at the interface: 
\begin{equation}
    A= B^\top B,
\end{equation}
where the matrix $B$ is a rectangular matrix, assembled from the phonon dynamical matrix \cite{li2025}. 
 Let \(n_w\) denote the row dimension. The exterior columns form \(B\in\mathbb R^{n_w\times N_{\mathcal E}}\), and the interface columns form \(L\in\mathbb R^{n_w\times N_I}\). The interface includes every core coordinate on a retained row.  The resulting Gram blocks and the deformation \(\bm w\in\mathbb R^{n_w}\) are summarized here
\begin{equation}
 B^\top B=A,\qquad B^\top L=C,\qquad
 \bm w=B\bm y+L\bm u_\Gamma .
 \label{eq:compatible-gram}
\end{equation}
The energy \(\|\bm w\|^2/2\) on these rows includes both ends of every crossing interaction. Its mixed term is \(\bm y^\top B^\top L\bm u_\Gamma\), which recovers the coupling term in \eqref{eq:energy}. If \(\bm e_a\) selects one interface coordinate, the column \(\bm\ell_a=L\bm e_a\) has support only on the rows touching that coordinate. The energy coordinate \(\bm w\) itself can extend throughout the crystal. For a prescribed trace the traction component \(a\) can now be expressed as 
\begin{equation}
 (\bm\tau)_a=(D_\times\bm u_\Gamma)_a+\bm\ell_a^\top B\bm y
 =[(D_\times-L^\top L)\bm u_\Gamma]_a+\bm\ell_a^\top\bm w .
 \label{eq:strain-force}
\end{equation}
The core knows the first term on the right. The second is one signed projection of the evolving deformation. Section~\ref{sec:algorithm} estimates this projection on the same Verlet steps as the core.

The force constants determine \(B\) through the factorization and supply \(C\) directly. A displacement probe \(C\bm e_a\) measures \((C^\top\bm y)_a\) directly, and Section~\ref{sec:precompute} uses these same columns to measure the reusable boundary response. Here we summarize the notations in the table below, together with some quantum primitives.  

\begin{center}
\begin{minipage}{.97\linewidth}
\small
\textbf{Notation and quantum operations}\smallskip

\begin{tabular}{@{}p{.25\linewidth}@{\hspace{.02\linewidth}}p{.73\linewidth}@{}}
\toprule
\(N_{\mathcal C}, N_{\mathcal E}, N_I\) & Numbers of scalar core, exterior and interface coordinates.\\
\(N_{\rm pair}\) & Number of distinct loading and readout probe pairs in the lag tables.\\
\(A,C,D\) & Exterior, exterior--interface and interface stiffness blocks in \eqref{eq:energy}.\\
\(B,L\) & Exterior and interface columns of the energy factor in \eqref{eq:compatible-gram}.\\
\(D_\times,\bm\tau\) & Core-end stiffness of the cut sum and its nodal holding traction.\\
\(F(t),\eta\) & Normalized exterior impulse response in \eqref{eq:lag-probe} and absolute error allowance for each unit-probe overlap.\\
\midrule
Block encoding & A unitary with signal block \(B/\alpha\) or \(H/\alpha\). The known scale \(\alpha\) bounds the highest exterior frequency \(\Omega\).\\
Generator query & One use of the supplied encoding or its inverse, controlled as needed.\\
Clean flags & Auxiliary qubits in their designated zero state select the signal block.\\
QSP / QSVT & Quantum signal processing and quantum singular value transformation apply the Verlet polynomials to an encoded operator.\\
LCU & A linear combination of unitaries adds the driven responses coherently.\\
QME & Quantum mean estimation resolves a signed overlap from interference.\\
\bottomrule
\end{tabular}
\end{minipage}
\end{center}

\section{Concurrent quantum--classical coupling}\label{sec:algorithm}
In a concurrent calculation, the classical core supplies a new interface displacement after every drift. The quantum calculation must return the signed traction exerted by this particular exterior field in time for the next kick (Fig.~\ref{fig:partition}c). We use the same Verlet step in both regions. This choice specifies which displacements drive the exterior and which quantum observable completes the classical step.

\subsection{One displacement and one traction per step}\label{sec:online}
We set \[X=I-\frac{\Delta t^2}2 A, \] and denote the initial exterior displacement by \(\bm y_0=\bm y(0)\). Eliminating the exterior velocities from~\eqref{eq:full-verlet} gives the driven position recurrence
\begin{align}
 \bm y_1&=X\bm y_0+\Delta t\,\dot{\bm y}(0)-\tfrac12(\Delta t)^2 C\bm u_{\Gamma,0},\nonumber\\
 \bm y_{n+1}&=2X\bm y_n-\bm y_{n-1}-(\Delta t)^2 C\bm u_{\Gamma,n},\qquad n\ge1.
 \label{eq:verletbath}
\end{align}

The last displacement needed to compute \(\bm y_{n+1}\) is \(\bm u_{\Gamma,n}\). The new displacement \(\bm u_{\Gamma,n+1}\) enters the instantaneous part of \(\bm\tau_{n+1}\). This ordering lets the core drift before the measurement. At the initial launch, the relative traction is \(D_\times\bm u_{\Gamma,0}+C^\top\bm y_0\). For an initially undisplaced exterior, the core evaluates it directly. For nonzero initial displacement, the term \(C^\top\bm y_0\) is supplied or measured within the traction tolerance.

\begin{center}\begin{minipage}{.94\linewidth}\hrule\smallskip
\refstepcounter{qmalg}\label{alg:march}\textbf{Algorithm \theqmalg. Concurrent exchange with matched Verlet}\smallskip
\begin{enumerate}[label=\arabic*.,ref=\arabic*,leftmargin=1.6em,itemsep=2pt,topsep=2pt]
\item\label{step:drift} \textbf{Core drift.} With acceleration \(-\nabla W(\bm q_n)-R^\top[(D-D_\times)\bm u_{\Gamma,n}+\bm\tau_n]\) from \eqref{eq:motion}, the first half kick and drift in \eqref{eq:full-verlet} give \(\bm q_m\), \(m=n+1\).
\item \textbf{Quantum exterior and force readout.}
\begin{enumerate}[label=\arabic{enumi}\alph*.,ref=\arabic{enumi}\alph*,leftmargin=2.2em,itemsep=1pt,topsep=2pt]
\item\label{step:load} The core sends \(\bm u_{\Gamma,m}=R\bm q_m\). For the history in \eqref{eq:live-strain} the controller stores \(L\Delta\bm u_{\Gamma,j}\), where \(\Delta\bm u_{\Gamma,j}:=\bm u_{\Gamma,j+1}-\bm u_{\Gamma,j}\) and the new step has \(j=m-1\).
\item\label{step:prepare} Quantum signal processing applies the Verlet polynomials in \eqref{eq:live-strain}. A linear combination of unitaries prepares their sum with signal \(\bm w_m/S_m\) and scale \(S_m\) from \eqref{eq:strain-scale}, when \(S_m>0\).
\item\label{step:read} For each required interface coordinate, quantum mean estimation supplies the signed ancilla mean in \eqref{eq:live-force}. The core rescales it and adds the local term in that equation to obtain \((\bm\tau_m)_a\). A zero signal scale or probe leaves only the local term.
\end{enumerate}
\item\label{step:kick} \textbf{Core kick.} With the new traction in \eqref{eq:motion}, the final half kick in \eqref{eq:full-verlet} completes the step. This acceleration also starts the next step.
\end{enumerate}\hrule\end{minipage}\end{center}
With exactly computed tractions and displacements, both kicks and the drift agree with full-scale Verlet by induction. One exchange per step suffices even when a reflected wave returns to the defect. Eliminating~\eqref{eq:verletbath} gives a discrete Dirichlet-to-Neumann (DtN) map with an instantaneous term, an incident force and a causal memory of the accepted displacements. E and Huang derived a time-discrete reflectionless boundary for a one-dimensional lattice without an incoming field~\cite[Sec.~4.1]{ehuang2002}. Related classical constructions use matching conditions, surface Green functions and rational approximations~\cite{li2006matching,monti2021,wu2018}. Here we allow general cut geometries and prescribed initial conditions in the exterior. 

\subsection{Preparing the driven field and measuring its traction}\label{sec:live-implementation}
\paragraph{Steps~\ref{step:drift} and~\ref{step:load}. The classical input}
A selected block of a unitary circuit has singular values at most one. We therefore divide the physical factor \(B\) by a known frequency scale \(\alpha\ge\|B\|_2\) and embed \(B/\alpha\) between designated input and output ancilla states. We use the compatible real factors \(A=B^\top B,\ C=B^\top L\) in~\eqref{eq:compatible-gram}. The strain on their retained rows is \(\bm w_m=B\bm y_m+L\bm u_{\Gamma,m}\). After the drift, the controller applies \(L\) to the new interface increment \(\Delta\bm u_{\Gamma,m-1}\) from Step~\ref{step:load}. It updates the circuit that selects a source and its weight from the stored history. Only this source data changes with the core motion. The block encoding of the lattice factor and the interface probes are fixed.

\paragraph{Step~\ref{step:prepare}. Quantum simulation using signal processing and superposition} A remarkable observation is that
the three-term Verlet recurrence leads to the Chebyshev propagators. Specifically, we introduce the standard Chebyshev polynomials as
\begin{equation}
    T_k(\cos\theta)=\cos(k\theta), U_k(\cos\theta)=\frac{\sin((k+1)\theta)}{\sin\theta}
\end{equation}
 with endpoint values by continuity, and \(U_{-1}=0\). 
 
 With \(X'=I-(\Delta t)^2BB^\top/2\), the exact Verlet solution for \(m\ge1\) is
\begin{equation}
\begin{aligned}
 \bm w_m={}&T_m(X')\bm w_0+\Delta t\,U_{m-1}(X')B\dot{\bm y}(0)\\
 &+\sum_{j=0}^{m-1}(U_{m-1-j}-U_{m-2-j})(X')\,L\Delta\bm u_{\Gamma,j}.
\end{aligned}\label{eq:live-strain}
\end{equation}
The first two terms carry the initial strain and velocity. Each accepted interface increment contributes once to the sum. The newest increment has coefficient \(I\), accounting for the new core end of the crossing interactions even though \(\bm y_m\) already follows from the preceding loads.

Quantum signal processing (QSP) applies these polynomials to the Hermitian generator \(H/\alpha\) in~\eqref{eq:generator} on the quantum circuit. Equivalently, quantum singular value transformation acts on the singular values of the rectangular factor \(B/\alpha\)~\cite{gilyen2019,low2019}. For an encoded singular value \(z\) the step argument is \(x=1-2c^2z^2\), where \(c=\Delta t\,\alpha/2<1\). The initial-strain and increment polynomials are even in \(z\). The velocity polynomial \(2cz\,U_{m-1}(x)\) is odd and takes a site velocity to strain. All three have degree at most \(2m\).

Each operator in \eqref{eq:live-strain} is implemented, after scaling, as a block of a unitary circuit and 
a linear combination of unitaries (LCU) adds the resulting signals~\cite{childswiebe2012}. The \(\ell^1\) norm of the coefficients sets the LCU preparation scale. With unit preparations of the nonzero sources, dividing the velocity and increment polynomials by \(b_{\Delta t}=(1-c^2)^{-1/2}\) gives the known LCU coefficient sum
\begin{equation}
 S_0=\|\bm w_0\|,\qquad
 S_m=\|\bm w_0\|+b_{\Delta t}
 \left(\|\dot{\bm y}(0)\|+\sum_{j=0}^{m-1}\|L\Delta\bm u_{\Gamma,j}\|\right).
 \label{eq:strain-scale}
\end{equation}
For a source supplied as a subnormalized block we use its preparation scale in this sum. Appendix~\ref{app:strain-increments} derives the polynomials and their scaling.

\begin{figure}[!htb]\centering\includegraphics[width=.97\linewidth]{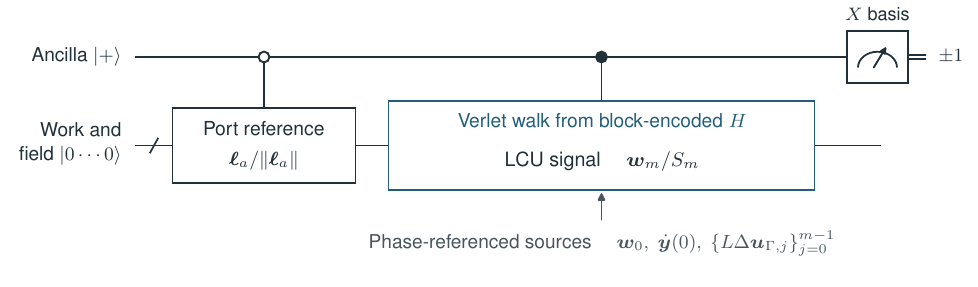}
\caption{Signed force readout in Algorithm~\ref{alg:march}. One branch prepares the interface probe and the other prepares the strain signal from the initial fields and accepted interface increments. The clean flags of the probe select the physical strain component. The ancilla interference mean gives the force through~\eqref{eq:live-force}.}\label{fig:circuit}
\end{figure}

\paragraph{Steps~\ref{step:read} and~\ref{step:kick}. Quantum mean estimation to extract the traction $\bm \tau$} 
In Fig.~\ref{fig:circuit}, we show how the circuit implementation of \eqref{eq:live-strain} is combined with an ancilla interference measurement to determine the traction. In particular,  an ancilla in the state \(|+\rangle\) coherently selects between preparing the interface probe \(|\mathrm{ref}\rangle\) and the driven field \(|\psi\rangle\). An \(X\)-basis measurement of this ancilla (a Hadamard gate followed by a computational-basis measurement) has signed mean \(\langle X_{\rm anc}\rangle=\operatorname{Re}\langle\mathrm{ref}|\psi\rangle\). Here \(|\psi_m\rangle\) is the normalized output of Step~\ref{step:prepare}, and its component with all signal flags zero is the vector \(\bm w_m/S_m\). The probe \(|\mathrm{ref}_a\rangle\) has clean flags and strain amplitudes \(\bm\ell_a/\|\bm\ell_a\|\), where \(\bm\ell_a=L\bm e_a\). The signed readout and physical traction are therefore
\begin{equation}
 \langle X_{\rm anc}\rangle=\frac{\bm\ell_a^\top\bm w_m}{\|\bm\ell_a\|S_m},
 \qquad
 (\bm\tau_m)_a=[(D_\times-L^\top L)\bm u_{\Gamma,m}]_a+
             \|\bm\ell_a\|S_m\langle X_{\rm anc}\rangle .
 \label{eq:live-force}
\end{equation}
Coherent quantum mean estimation (QME) estimates the probability \((1+\langle X_{\rm anc}\rangle)/2\) of a positive ancilla outcome, using the preparation, its inverse and reflections~\cite{brassard2002,montanaro2015}.

For a nonzero strain and probe, the joint probability of preparing the physical strain and projecting it onto the probe is \(|\langle\mathrm{ref}_a|\psi_m\rangle|^2\). Its factors are \(\|\bm w_m\|^2/S_m^2\) for strain preparation and \(|\bm\ell_a^\top\bm w_m|^2/(\|\bm\ell_a\|^2\|\bm w_m\|^2)\) for projection onto the probe conditional on that preparation. The latter can be regarded as the fraction of the elastic strain energy in the probe direction. A localized strain can give high contrast if it projects strongly onto that direction and the LCU scale is comparable to the strain norm.  In Appendix~\ref{sec:readout-prob} we give sufficient conditions for relative accuracy. For absolute component-force tolerance \(\varepsilon_\tau\), QME uses \(O(1+\|\bm\ell_a\|S_m/\varepsilon_\tau)\) preparations at fixed confidence, including near a force zero.

Each use of the preparation rebuilds the signal $\bm w_m$ at \(t_m\) from the accepted history. Thus the individual traction estimates and subsequent steps can use fresh copies. At a fixed margin \(c<1\) a replay takes \(O(m)\) factor queries up to precision logarithms. Loading the \(m\) local sources and their weights also costs \(O(m)\) operations up to addressing and precision factors for a fixed interface. The total number of factor queries is bounded by
\begin{equation}
 \begin{aligned}Q_{\rm online}&=\widetilde O\!\left(
 \sum_{m=1}^{N_T}m\sum_{a=1}^{N_I}
 \left[1+\|\bm\ell_a\|S_m/\varepsilon_\tau\right]\right)
 \\&=\widetilde O\!\left(N_I N_T^2[1+F_{\max}/\varepsilon_\tau]\right),
 \qquad F_{\max}=\max_{m,a}\|\bm\ell_a\|S_m .\end{aligned}
 \label{eq:direct-cost}
\end{equation}
The quadratic step count comes from replaying successively longer histories. The factor \(F_{\max}/\varepsilon_\tau\) converts the worst preparation scale to absolute force accuracy. Wave spreading can further weaken the relative signal at an atom at the interface. Launch and probe preparations are charged on each use, with confidence amplification included in the logarithmic factors. A displacement probe also permits live exchange with independently supplied crossing blocks (Appendix~\ref{sec:direct}). In Section~\ref{sec:precompute} we measure the response to prescribed loads offline, leaving the evolving interface history to the classical core.

\section{Quantum precomputation of the exterior response}\label{sec:precompute}
\subsection{Encoding harmonic dynamics in a Schr\"odinger equation and the generalized boundary moments}\label{sec:ct}
The waves launched by an interface impulse depend only on the material, the partition of the system, and the outer boundary. This makes an offline--online calculation possible. We compute the impulse response of the harmonic exterior offline from the exterior propagator and use it online for each successive core displacement. To construct the propagator we use an energy factor \(A=B^\dagger B\) in \eqref{eq:compatible-gram}. The adjoint accommodates complex factors in a Fourier representation and equals the transpose for the real factor of Section~\ref{sec:portpower}. 

On the site and strain spaces \(\mathbb C^{N_{\mathcal E}}\oplus\mathbb C^{n_w}\), the Hermitian dilation of \(B\) and the physical state take the form
\begin{equation}
 H=\begin{pmatrix}0&B^\dagger\\ B&0\end{pmatrix},\qquad
 H^2=\operatorname{diag}(A,BB^\dagger),\qquad
 \bm\phi(t)=\binom{\dot{\bm y}(t)}{-iB\bm y(t)} .
 \label{eq:generator}
\end{equation}
As input models, selecting \(B\) or \(B^\dagger\)  gives a block encoding of \(H/\alpha\) from one of \(B/\alpha\), where \(\alpha\ge\|B\|_2\) is a known frequency scale. 

Differentiating the two components using \eqref{eq:motion} gives the driven Schr\"odinger equation and its solution,
\begin{align}
 i\dot{\bm\phi}(t)&=H\bm\phi(t)-i\binom{C\bm u_\Gamma(t)}{0},
 \label{eq:schrodinger}\\
 \bm\phi(t)&=e^{-itH}\bm\phi(0)-
       \int_0^t e^{-i(t-s)H}\binom{C\bm u_\Gamma(s)}{0}\,ds .
 \label{eq:duhamel}
\end{align}
The vector \(\bm\phi\) carries physical amplitudes. In free motion its squared norm is twice the harmonic energy. During forcing, its derivative satisfies \(d\|\bm\phi\|^2/dt=-2\dot{\bm y}^\top C\bm u_\Gamma\). An initially undisplaced exterior with zero velocity has \(\bm\phi(0)=0\). 

For the circuit we zero-pad the two sectors in the dilation \eqref{eq:generator} to a common address dimension, with sector label \(|0\rangle\) for sites and \(|1\rangle\) for strain. We also write \(|b,\bm z\rangle=|b\rangle\otimes|\bm z\rangle\) for a padded vector in sector \(b\).  We define
\begin{equation}
\begin{aligned}
 F(t)&=(\langle0|\otimes I)\operatorname{sinc}(tH)(|0\rangle\otimes I),\\
 f(t)&=\bm p^\dagger F(t)\bm z
      =\langle0,\bm p|\operatorname{sinc}(tH)|0,\bm z\rangle
      =\int_0^1\langle0,\bm p|e^{-irtH}|0,\bm z\rangle\,dr .
\end{aligned}
 \label{eq:lag-probe}
\end{equation}
We call \(f(t)\) the \emph{generalized response moment} at lag \(t\) for unit site probes \(\bm p,\bm z\in\mathbb C^{N_{\mathcal E}}\). We used \(\operatorname{sinc}(x)=\sin(x)/x\) with \(\operatorname{sinc}(0)=1\). This came up because on a normal mode of stiffness \(\omega^2\), the velocity response is \(\cos(\omega t)\) and the displacement response is \(t\operatorname{sinc}(\omega t)\). A zero-frequency velocity impulse therefore gives displacement proportional to \(t\). 

Equations~\eqref{eq:generator} and~\eqref{eq:lag-probe} express the moment entirely through the encoded generator. In particular, 
integrating \eqref{eq:duhamel} and adding the initial displacement gives the traction that closes the core equations,
\begin{equation}
\begin{aligned}
 \bm\tau(t) &=D_\times\bm u_\Gamma(t)+\bm\tau^{\rm inc}(t)
     -\int_0^t(t-s)\,C^\top F(t-s)C\bm u_\Gamma(s)\,ds,\\
 \bm\tau^{\rm inc}(t)&=C^\top(\langle0|\otimes I)e^{-itH}|0,\bm y_0\rangle+
                        t C^\top F(t)\dot{\bm y}(0).
\end{aligned}
\label{eq:ct-traction}
\end{equation}
The first two terms give the instantaneous holding force and the incident force. In structural dynamics, \(tF(t)\) is the displacement Green operator of the clamped exterior. Its \(N_I\times N_I\) interface projection \(tC^\top F(t)C\) enters the traction memory with a minus sign because the core loads the exterior through \(-C\bm u_\Gamma\). In \eqref{eq:lag-probe}, we choose \(\bm p\) and \(\bm z\) along \(C\bm e_a\) and \(C\bm e_b\), respectively. Multiplying their unit-probe overlap by the two column norms gives entry \((a,b)\).

The same overlap supplies all the time-dependent exterior data in \eqref{eq:ct-traction}. For example, for the velocity term we keep the same readout \(\bm p\) and choose \(\bm z\) along \(\dot{\bm y}(0)\). For the displacement term we set \(\bm k_0=A\bm y_0\) and use the identity
\begin{equation*}
 (\langle0|\otimes I)e^{-itH}|0,\bm y_0\rangle
    =\bm y_0-\int_0^t sF(s)\bm k_0\,ds .
 \tag{\ref*{eq:ct-traction}a}\label{eq:incident-position}
\end{equation*}
Thus the displacement contribution uses the same readout along \(C\bm e_a\), an input along \(\bm k_0\), and a time integral of the rescaled moments. A zero input or readout column contributes zero. The static offset \(C^\top\bm y_0\) is supplied with the launch or measured separately at lag zero with input along \(\bm y_0\). 

\begin{center}\begin{minipage}{.94\linewidth}\hrule\smallskip
\refstepcounter{qmalg}\label{alg:reusable}\textbf{Algorithm \theqmalg. Quantum response setup and classical defect dynamics}\smallskip
\begin{enumerate}[label=\arabic*.,leftmargin=1.6em,itemsep=2pt,topsep=2pt]
\item \textbf{Quantum setup.} For each nonzero probe pair, the circuit estimates the normalized lags required by \eqref{eq:shifted-both}. Equations~\eqref{eq:lag-probe} and \eqref{eq:global-error} give the continuous-time overlap and its interpolation accuracy. Proposition~\ref{prop:dyadic} also gives a shared-interval construction for either continuous or matched-Verlet lags. The solver stores the rescaled interface and incident responses in \eqref{eq:shifted-both}. For the continuous choice, \eqref{eq:incident-position} supplies the initial-displacement contribution.
\item \textbf{Classical force update.} At launch, \(\bm\tau_0=D_\times\bm u_{\Gamma,0}+C^\top\bm y_0\) supplies the first half kick. After each drift in \eqref{eq:full-verlet}, the core sets \(\bm u_{\Gamma,n}=R\bm q_n\) and evaluates \(\bm\tau_n\) from \eqref{eq:shifted-both}. Earlier displacements enter the convolution. The current displacement enters the instantaneous term, since the displacement impulse vanishes at lag zero.
\item \textbf{Classical kick.} The local force and \(\bm\tau_n\) give the core acceleration in \eqref{eq:motion} for the final half kick in \eqref{eq:full-verlet}. Every subsequent drift and kick reuses the material response \(C^\top F_\ell C\).
\end{enumerate}\hrule\end{minipage}\end{center}

For the interface defined in Section~\ref{sec:partition}, a dense reciprocal boundary table needs \(N_I(N_I+1)/2\) pairs. Each nonzero incident vector adds at most \(N_I\), so \(N_{\rm pair}\le N_I(N_I+1)/2+2N_I\). A fixed number of loading directions gives at most linear growth in \(N_I\). 

Let us now turn to the dynamics of the core atoms. Notice that the Verlet kicks in \eqref{eq:full-verlet} require the traction at integer times \(t_n=n\Delta t\). Writing \(F_\ell\) for the exterior response operator at lag \(t_\ell\), we compute that traction from
\begin{equation}
\begin{aligned}
 \bm\tau_n=D_\times\bm u_{\Gamma,n}+C^\top\bm y_0
   &+t_n C^\top F_n\dot{\bm y}(0)\\
   &-\Delta t\sum_{j=0}^{n-1}\nu_j\,t_{n-j}\,
        C^\top F_{n-j}(C\bm u_{\Gamma,j}+\bm k_0).
\end{aligned}\label{eq:shifted-both}
\end{equation}
For \(F_\ell=F(t_\ell)\), this is the composite trapezoidal rule applied to the continuous history, with \(\nu_0=1/2\) and \(\nu_j=1\) for \(0<j<n\). The omitted endpoint \(j=n\) has zero impulse. For smooth histories, higher-order rules such as composite Simpson quadrature are also possible. The analysis below uses the trapezoidal weights. For a unit velocity impulse the position recurrence \eqref{eq:verletbath} gives \(\Delta t\,U_{\ell-1}(X)\) at step \(\ell\). Since \(t_\ell=\ell\Delta t\), the choice \(F_\ell=P_\ell=U_{\ell-1}(X)/\ell\) for \(\ell\ge1\) gives the exact matched-Verlet history, including the initial half kick. We set \(F_0=I\), since the velocity term at launch is multiplied by \(t_0=0\). In both cases the core uses the interface matrix \(C^\top F_\ell C\) in the same classical update.

\paragraph{Several defect regions}
For disjoint core regions sharing one exterior, we order the columns as \(C=[C_1\ C_2\ \cdots]\). Motion at interface \(j\) contributes to the force history at \(i\) through the block \(-sC_i^\top F(s)C_j\) at lag \(s\). The reverse block is its transpose. Applying Algorithm~\ref{alg:reusable} to the concatenated traces therefore retains repeated scattering between defects in the chosen harmonic-exterior model. The pair count uses the total interface size.

\subsection{Global polynomial interpolation at Chebyshev times}\label{sec:timesamples}
Rather than computing the generalized moments at all the time steps, we may pre-compute them at fewer steps and then apply interpolation later. 
For each unit probe pair, we propose the quantum circuit to estimate the normalized response \(f\) in \eqref{eq:lag-probe} at \(M+1\) Chebyshev--Lobatto times,
\[
 s_j=\tfrac T2\bigl(1-\cos(j\pi/M)\bigr),\qquad j=0,\ldots,M,\quad M\ge1 .
\]
At each node the circuit averages the simulated propagator up to \(s_j\) and estimates the signed overlap by coherent amplitude estimation. The value at zero is the overlap of the two probes and is one for identical probes. We fit one polynomial in physical lag time to the measured values and evaluate it at the equally spaced lags \(t_n=n\Delta t\) requested by the core.

Polynomial interpolation with noisy data requires the control of bias and variance to ensure good accuracy. 
Two properties of the harmonic response govern the required number of measurements. On a normal mode of frequency \(\omega\), the normalized response is \(\operatorname{sinc}(\omega t)\). Its first oscillation can occupy a small fraction of a long observation interval, where the Chebyshev nodes cluster near the initial time. Subsequent oscillations decrease in amplitude. At a fixed horizon the bounded frequency spectrum also makes the response analytic, allowing more rapid convergence as the degree increases.

For each pair $(\bm p, \bm z)$ in \eqref{eq:lag-probe}, we use the lag tolerance \(\eta\) to bound the absolute error in its dimensionless response \eqref{eq:lag-probe} at every requested time. The following theorem and Proposition~\ref{prop:dyadic} summarize the quantum overhead of constructing a response table at that accuracy, with the standard quantum input model of controlled block encoding of \(H/\alpha\), or of
\(B/\alpha\) and its adjoint, together with coherent preparations and
their inverses for \(N_{\rm pair}\) pairs of unit probes with known
physical scales.

\begin{theorem}[Quantum estimation at Chebyshev lag times]\label{thm:time-moments}
For each pair $(\bm p, \bm z)$ in \eqref{eq:lag-probe}, let \(\widehat f_M\) be the polynomial of degree at most
\(M\) interpolating the measured values of the response \(f\) in
\eqref{eq:lag-probe} at the \(M+1\) Chebyshev--Lobatto times on
\([0,T]\). Let \(\Lambda_M\) be the Lebesgue constant of these nodes
and \(\Omega=\|H\|_2\) the highest exterior angular frequency.
If every measured value has absolute error at most \(\sigma\), then
the interpolant satisfies
\begin{equation}
 \Big\|f-\widehat f_M\Big\|_{\infty,[0,T]}
 \le \Lambda_M\sigma+
 \min\!\left\{
  \frac{9\pi^2(1+\Lambda_M)\Omega T}{32(M+1)^2},\quad
  4e^{3\Omega T/8}2^{-M},\quad
  \frac{4(\Omega T/4)^{M+1}}{(M+2)!}
 \right\}.
 \label{eq:global-error}
\end{equation}
For a target lag accuracy \(0<\eta<1\), we choose the smallest
\(M\ge1\) for which the interpolation term (the minimum above) is at
most \(\eta/2\), and set \(\sigma=\eta/(2\Lambda_M)\).

For \(0<\delta<1\), the quantum algorithm achieves uniform error at
most \(\eta\) for all probe pairs simultaneously with probability at
least \(1-\delta\), using
\begin{equation}
 \widetilde O\!\left(
 \frac{N_{\rm pair}(1+\alpha T)}{\eta}
 \left[1+\min\!\left\{\sqrt{\Omega T/\eta},\,
                  \Omega T+\log(1/\eta)\right\}\right]\right)
 \label{eq:time-sample-cost}
\end{equation}
controlled generator queries. 
\end{theorem}

At fixed normalized tolerance, the first bias bound in \eqref{eq:global-error} permits a degree of order \(\sqrt{\Omega T}\) up to logarithms. For fixed \(0<\eta<1/2\), a single harmonic mode already requires degree at least proportional to \(\sqrt{\Omega T}\) for uniform polynomial approximation as the horizon grows (Appendix~\ref{app:time-estimator}). At fixed \(\Omega T\) the third bound gives factorial decay with degree. These statements concern the normalized lag. Section~\ref{sec:fixedbudget} relates its accuracy to the displacement of the defect.

Overall the smoothness of the moments controls the interpolation bias, and interpolation weights determine how measurement errors pass to each reconstructed lag. This distinction also arises in quantum zero-noise extrapolation~\cite{mohammadipour2025zne}. Here all lags lie inside the sampled interval. Independent unbiased node estimates of variance at most \(v\) give variance at most \(2v\) at each lag, independently of degree (Appendix~\ref{app:time-estimator}). The trajectory bound in the following sections uses the simultaneous nodal error event in \eqref{eq:global-error}, which also covers biased estimates and correlations between reconstructed lags.

\subsection{Shared interval averages for the complete lag table}\label{sec:dyadic}
The core's dynamics simulation needs a force at every time step, but the normalized displacement responses at successive lags contain much of the same exterior motion. For a pair of site probes, the lag \(f(t_n)\) is the mean velocity response over \([0,t_n]\). We can assemble this mean from shorter integrals. For time steps \(i<j\), write
\begin{equation}\label{interval}
g_{i,j}=\frac{1}{(j-i)\Delta t}
       \int_{t_i}^{t_j}\langle0,\bm p|e^{-isH}|0,\bm z\rangle\,ds .
\end{equation}

The circuit estimates \(g_{i,j}\) directly as a signed probe overlap with a bounded polynomial of \(H\) (Appendix~\ref{app:dyadic-proof}). We can use aligned intervals of power-of-two length so that one measurement can serve many lags.

\smallskip
\noindent\begin{minipage}[c]{.425\linewidth}
\centering\includegraphics[width=.98\linewidth]{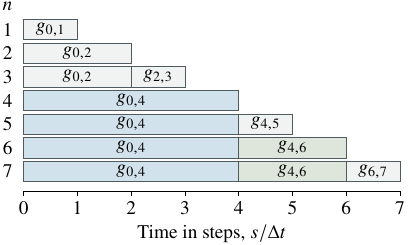}
\end{minipage}\hfill
\begin{minipage}[c]{.54\linewidth}\small
Each row partitions \([0,t_n]\). A block's width divided by the row length is its weight in \(f(t_n)\). For example,
\[
  f(t_7)=\tfrac47 g_{0,4}+\tfrac27 g_{4,6}+\tfrac17 g_{6,7}.
\]
The first mean serves rows 4 through 7 and the second serves rows 6 and 7. An error in the last mean contributes only one seventh of its value to the seventh lag.
\end{minipage}
\par\smallskip

There is one candidate block ending at each \(n\). Its width in time steps, \(\kappa_n\), is the largest power of two dividing \(n\). With an accumulator \(h_0=0\), one classical pass gives
\[
 h_n=h_{n-\kappa_n}+\kappa_n\widehat g_{n-\kappa_n,n},\qquad
 \widehat f(t_n)=h_n/n,\qquad 1\le n\le N_T .
\]
Each lag uses at most \(J=\lceil\log_2(2N_T)\rceil\) blocks. A block of length \(\kappa\) ending at \(j\) has weight \(\kappa/n\le\kappa/j\) wherever it occurs. We can therefore estimate its mean to error \(\eta_j=\eta j/(J\kappa)\). The error in any one lag sums to at most \(\eta\). A short interval late in the history can have a large allowance. When \(\eta_j\ge1\), zero already meets it since \(|g_{j-\kappa,j}|\le1\).

The quantum saving comes from this variation in precision as well as from reuse. To see this, we notice that reaching step \(j\) costs \(\widetilde O(1+\alpha j\Delta t)\) generator queries per circuit use. For a block that requires measurement, coherent mean estimation takes \(\widetilde O(1/\eta_j)\) uses. At fixed width \(\kappa\), the factor \(j\) in the propagation cost cancels against the \(1/j\) in the required number of uses. Only endpoints \(j=k\kappa\) with \(k<J/\eta\) can require measurement, giving \(O(\min\{N_T,J^2/\eta\})\) measured means over all widths per pair. 

A related hierarchy has been used widely to speed up the time-marching cost. Hairer, Lubich and Schlichte grouped past loads into blocks for FFT convolution with a known kernel~\cite{hairer1985volterra}. Kaye and Strand extended fast history summation to a kernel determined by the evolving solution~\cite{kaye2023dyson}. Here the interval measurements  \eqref{interval} construct the moments separately from the defect dynamics of $\bm q(t)$. The classical march then applies the known kernel to the displacements as they become available.

The following proposition summarizes the query cost of the interval construction. 
\begin{proposition}[Query cost of all normalized lags]\label{prop:dyadic}
Let \(t_n=n\Delta t,\ T=N_T\Delta t\), and suppose that a controlled block encoding of \(H/\alpha\), coherent preparations of the \(N_{\rm pair}\) pairs of unit site probes and their inverses are available. All entries \(f(t_n)=\langle0,\bm p|\operatorname{sinc}(t_nH)|0,\bm z\rangle\) can be estimated to absolute error \(\eta\in(0,1)\), simultaneously with probability at least \(1-\delta\), using
\[
   \widetilde O\!\left(\frac{N_{\rm pair}(1+\alpha T)}{\eta}\right)
   \quad\text{quantum queries},\qquad
   O(N_{\rm pair}N_T)\quad\text{classical operations to form the table}.
\]
The same bounds hold for the matched-Verlet lags \(P_n=U_{n-1}(I-(\Delta t)^2A/2)/n\) when \(\alpha\Delta t/2\le1\). The query count suppresses logarithmic precision, simulation and joint-confidence factors.
\end{proposition}

 The Chebyshev polynomial of Theorem~\ref{thm:time-moments} gives a continuous interpolant, with factorial decay of the bias at fixed \(\Omega T\). In contrast, when only the Verlet time steps are needed, sharing intervals removes the node-count factor in \eqref{eq:time-sample-cost}. For an unknown encoded exterior, even one late continuous lag can require order \(\alpha T/\eta\) queries at fixed success probability above one half, for \(\alpha T\ge2\pi\) and sufficiently small \(\eta\) (Appendix~\ref{app:dyadic-proof}).

\subsection{Stiffness moments for a convolution-free Verlet march}
\label{sec:stiffness-moments}

The classical force update in Algorithm~\ref{alg:reusable} uses the projected
displacement response \(C^\top F_\ell C\) in~\eqref{eq:shifted-both}.
We can construct this response from polynomials of the exterior stiffness
\(A\). For matched Verlet, this representation lets the core advance a set of
interface variables without storing past displacements. We use it in the
matched-Verlet benchmarks of Sections~\ref{sec:incoming-test}
and~\ref{sec:matched-work}.

For the unit site probes \(\bm p,\bm z\) of~\eqref{eq:lag-probe}, define
\begin{subequations}
\begin{equation}
 Y=\frac{2A}{\alpha^2}-I,\qquad
 \mu_k(\bm p,\bm z)=\bm p^\dagger T_k(Y)\bm z,\qquad
 M_k=C^\top T_k(Y)C,\qquad 0\leq k\leq K .
 \label{eq:sm-moments}
\end{equation}
The scale \(\alpha>0\) satisfies \(\|A\|_2\leq\alpha^2\), so the spectrum of
\(Y\) lies in \([-1,1]\). Measuring between normalized nonzero columns of \(C\)
and restoring their norms gives the entries of \(M_k\).
By~\eqref{eq:generator}, the site block of \(H^2\) is \(A\). The identity
\(T_{2k}(u)=T_k(2u^2-1)\) therefore makes \(T_k(Y)\) the site block of
\(T_{2k}(H/\alpha)\). An order-\(2k\) qubitized walk supplies the signal for
\(\mu_k(\bm p,\bm z)\), which signed mean estimation resolves to the required
precision. For explicitly supplied local columns, \(M_0=C^\top C\) is
directly available.

Let \(L_j\) be the cardinal polynomial of the stiffness node
\(y_j=\cos(j\pi/K)\), with \(0\leq j\leq K\) and \(K\geq1\). Set
\(\lambda_j=\alpha^2(1+y_j)/2\). Expanding \(L_j\) in Chebyshev polynomials
and combining the measured moments in~\eqref{eq:sm-moments} gives
an estimate \(\widehat G_j\) of each node matrix \(G_j\). For a scalar
stiffness function \(\psi\), we use
\begin{equation}
 G_j=C^\top L_j(Y)C,\qquad
 C^\top \psi(A)C \ \simeq\ \sum_{j=0}^K \psi(\lambda_j)\widehat G_j .
 \label{eq:sm-projection}
\end{equation}
With exact moments the right-hand side is the interface projection of the
degree-\(K\) stiffness interpolant of \(\psi\).

For matched Verlet, the exterior recurrence~\eqref{eq:verletbath} gives the
normalized lag \(P_\ell=U_{\ell-1}(X)/\ell\) in~\eqref{eq:shifted-both}.
At the stiffness node \(\lambda_j\), \(X=I-(\Delta t)^2A/2\) has scalar
counterpart \(\chi_j=1-(\Delta t)^2\lambda_j/2\). Assume
\(\alpha\Delta t/2\leq1\). If \(\bm y_0=\dot{\bm y}(0)=0\), the three-term
Verlet recurrence at each node advances one interface vector:
\begin{equation}
\begin{aligned}
 \bm\xi_{j,0}&=0,\qquad
 \bm\xi_{j,1}=\tfrac12(\Delta t)^2\bm u_{\Gamma,0},\\
 \bm\xi_{j,n+1}&=2\chi_j\bm\xi_{j,n}-\bm\xi_{j,n-1}
                  +(\Delta t)^2\bm u_{\Gamma,n},\qquad n\geq1 .
\end{aligned}
 \label{eq:sm-aux}
\end{equation}
\end{subequations}
Indeed, \(t_\ell P_\ell=\Delta t\,U_{\ell-1}(X)\), so at each node the
recurrence accumulates the past interface loads with the Verlet weights
in~\eqref{eq:shifted-both}, including \(\nu_0=1/2\). The delayed holding
force for the interpolated kernel is \(-\sum_{j=0}^K\widehat G_j\bm\xi_{j,n}\).
Adding \(D_\times\bm u_{\Gamma,n}\) gives the holding traction for this
launch. Appendix~\ref{app:stiffness-moments} gives the incident terms for
nonzero exterior initial data. For fixed \(K\), a dense interface requires
\(O((K+1)N_I^2)\) response storage and the same order of arithmetic per
step, in addition to the core force calculation. The two most recent
auxiliary states suffice.

\paragraph{Precision of the measured moments}
For the normalized continuous response \(F(t_\ell)\) of~\eqref{eq:lag-probe}
and the matched-Verlet response \(P_\ell\) in~\eqref{eq:shifted-both},
the degree-\(K\) stiffness interpolant has coefficients \(a_{\ell k}\)
satisfying \(\sum_{k=0}^K|a_{\ell k}|=1\) by~\eqref{eq:sm-coeff-mass}.
Consequently, simultaneous absolute errors at most \(\eta/2\) in the
unit-probe moments contribute at most \(\eta/2\) to every normalized lag.
A degree \(K=O(1+\alpha T+\log(1/\eta))\) suffices for the remaining
interpolation error uniformly up to time \(T\). For matched Verlet the lag
at step \(\ell\) has degree \(\ell-1\), so \(K=N_T-1\) already removes the
interpolation bias when \(N_T\geq2\). Under the access assumptions of
Proposition~\ref{prop:dyadic}, measuring order \(k\geq1\) to absolute error
\(\eta/2\) takes \(\widetilde O(k/\eta)\) generator queries per probe pair.
Summing over orders gives \(\widetilde O(N_{\rm pair}K^2/\eta)\) queries
with joint success probability \(1-\delta\), up to logarithmic confidence
factors. Here \(N_{\rm pair}\) includes any incident probes. The physical
gain in~\eqref{eq:total-gain} converts this lag tolerance to a
core-displacement error allowance. Shared intervals have the stronger
worst-case horizon bound in Proposition~\ref{prop:dyadic}. Stiffness
moments provide the convolution-free march~\eqref{eq:sm-aux} when a small
\(K\) resolves the responses of interest.

\section{Coupling accuracy and computational cost}\label{sec:coupling}
\subsection{Stability of the core trajectory under traction errors}\label{sec:robustness}
After eliminating the exterior, the core obeys a nonlinear Volterra equation with memory. Its stability can be studied by restoring the exterior variables and applying a standard continuous-dependence argument to the full Verlet recurrence. The perturbation forcing in this argument is the difference between the traction used by the core and the traction returned by the exact exterior driven by the same \emph{computed} interface trace. The instantaneous terms involving \(D_\times\) use the same core displacement, so their contribution to the force discrepancy cancels. An exterior readout error thus enters the core dynamics as an additive force. We denote the largest norm of this force by \(r_{\rm num}\), including any discrepancy between the force used and the force on the accepted trace. 

We set up this analysis so that the two trajectories start from the same positions and velocities. We write the model acceleration as \(-\mathsf K_{\rm lin}\bm{u}+\bm{f}_{\rm nl}(\bm{u})\), where \(0\preceq(\Delta t)^2 \mathsf K_{\rm lin}\preceq4I\) and \(L_{\rm nl}\) bounds the Lipschitz constant of \(\bm{f}_{\rm nl}\) in a tube enclosing the comparison. For the selected reference core displacement \(\bm{q}^{\rm ref}_n\), we prove in Appendix~\ref{app:error-cost} that 
\begin{equation}
 \max_{0\le n\le N_T}\|\widehat{\bm{q}}_n-\bm{q}^{\rm ref}_n\|_2
 \le E_{\Delta t}+S_x r_{\rm num},\qquad
 S_x=\frac{\cosh(T\sqrt{L_{\rm nl}})-1}{L_{\rm nl}} .
 \label{eq:traj}
\end{equation}
At \(L_{\rm nl}=0\), we use the limiting value \(S_x=T^2/2\). The Verlet error propagator gives \eqref{eq:traj}, and its discrete energy balance identifies the work done by a traction error on the velocity error (Appendix~\ref{app:error-cost}). A comparison with  full-scale Verlet on the prescribed model has \(E_{\Delta t}=0\), while comparison with a smooth continuous reference also incurs \(E_{\Delta t}=O((\Delta t)^2)\).

\subsection{Overall overhead at a prescribed accuracy}\label{sec:fixedbudget}
The lag tolerance \(\eta\) of Section~\ref{sec:timesamples} bounds the absolute error in each normalized probe overlap, uniformly over the lags used in \eqref{eq:shifted-both}. Proposition~\ref{prop:dyadic} uses the same criterion for matched Verlet. To express this dimensionless tolerance in terms of a defect displacement, we restore the units of each measured pair, add the time weights in the actual history rule, and propagate the resulting force error through the nonlinear core. With \(u_{\max}\) bounding \(\|R\widehat{\bm q}_n\|_2\) along the computed trajectory, we define
\begin{equation}
 B_T^{\rm tot}=S_x\!\Big[\tfrac12 T^2 u_{\max}\|C\|_F^2+
      \|C\|_F\big(T\|\dot{\bm y}(0)\|+\tfrac12 T^2\|\bm k_0\|\big)\Big].
 \label{eq:total-gain}
\end{equation}
This parameter allows us to combine the error and evaluate the overall cost. The following theorem gives the quantum cost of building the table and the classical cost of using it for one trajectory at the requested displacement accuracy.

\begin{theorem}[Cost of a core trajectory at prescribed accuracy]\label{thm:cost}
Given a core-displacement tolerance \(\varepsilon>0\), suppose the discretization and input errors for the chosen harmonic-exterior model leave at least \(\varepsilon/2\) for estimating the exterior response. Assume the stability bound \eqref{eq:traj}, the coherent access of Proposition~\ref{prop:dyadic}, and a supplied launch offset \(C^\top\bm y_0\) within the input error budget. For either the continuous or the matched-Verlet response, there is a lag tolerance \(\eta\) for which Algorithm~\ref{alg:reusable} achieves \(\max_{0\le n\le N_T}\|\widehat{\bm q}_n-\bm q_n^{\rm ref}\|\le\varepsilon\) with probability at least \(1-\delta\). With the shared intervals of Section~\ref{sec:dyadic} and bounded force-evaluation work per core coordinate, computing the trajectory through \(T\) requires
\begin{equation}
\begin{aligned}
 &\widetilde O\!\left(N_{\rm pair}(1+\alpha T)
       \max\{1,B_T^{\rm tot}/\varepsilon\}\right)
      &&\text{quantum queries for setup},\\
 &O\!\left(N_{\rm pair}N_T\right)
      &&\text{classical operations to form the table},\\
 &O\!\left(N_T N_{\mathcal C}+N_I^2 N_T\log^2(2N_T)\right)
      &&\text{classical operations to advance the core}.
\end{aligned}
\label{eq:main-cost}
\end{equation}

\end{theorem}

Appendix~\ref{app:cost-proof} chooses the normalized lag tolerance and proves the work bounds. Table~\ref{tab:simplified} compares both coupling algorithms with  full-scale Verlet.

\medskip 
\begin{table}[!htb]\centering\small
\caption{Work for one trajectory with supplied lattice and probe circuits. The query bounds omit logarithmic factors. The classical operation bounds assume a bounded stencil and include core force evaluations of cost \(C_F\). Each offline method reuses its material kernel across core laws and launches for the same cut and exterior. Each new incident field needs its own response.}\label{tab:simplified}
\begin{tabular}{@{}llll@{}}\toprule
Calculation & Generator queries & Classical march & \shortstack[l]{Reusable\\kernel}\\\midrule
 full-scale Verlet & \(0\) & \(N_T(N_{\mathcal E}+C_F)\) & ---\\
Concurrent Verlet & \(N_I N_T^2(1+F_{\max}/\varepsilon_\tau)\) & \(N_T(C_F+N_I^2)\) & ---\\
Chebyshev times & \((M+1)N_{\rm pair}(1+\alpha T)/\eta\) & \(N_T[C_F+N_I^2\log^2(2N_T)]\) & Yes\\
Shared intervals & \(N_{\rm pair}(1+\alpha T)/\eta\) & \(N_T[C_F+N_I^2\log^2(2N_T)]\) & Yes\\\bottomrule
\end{tabular}

\end{table}

In Table~\ref{tab:simplified}
the concurrent march counts the classical core and interface updates. Loading the past sources for each quantum replay is accounted for in Section~\ref{sec:live-implementation}.  For the global polynomial, \(M\) is the degree selected in Theorem~\ref{thm:time-moments} at lag tolerance \(\eta\). Measuring one node \(s_j\) to precision \(\sigma\) costs \(\widetilde O((1+\alpha s_j)/\sigma)\) generator calls per pair. Equation~\eqref{eq:time-sample-cost} sums these costs at the degree and precision required by \eqref{eq:global-error}. Once the values are known, direct evaluation costs \(O(M)\) classical operations per lag and pair. Fast multipoint evaluation can also form all the lags together. 

The last row measures the interval means of Section~\ref{sec:dyadic}. Theorem~\ref{thm:cost} charges their classical assembly and the subsequent core march separately. To convert any setup count to gates one needs the cost of a controlled encoding, the cost of each probe preparation and an arithmetic precision compatible with the trajectory tolerance. Appendix~\ref{app:time-estimator} gives the additional preparation cost when only a block encoding of \(C\) is supplied.

\paragraph{Dependence on the exterior size}
The gain over full-scale Verlet comes from computing the response at the
interface once and then advancing only the nonlinear core. For a
bounded interaction stencil, full-scale Verlet updates all
\(N_{\mathcal E}\) exterior coordinates at each step, at a classical
cost of \(\Theta(N_T(N_{\mathcal E}+N_{\mathcal C}))\). In
Theorem~\ref{thm:cost}, the quantum setup count is independent of
\(N_{\mathcal E}\) at fixed core, interface, horizon and displacement
tolerance, provided the encoding scale \(\alpha\) and the physical gain
\(B_T^{\rm tot}\) remain bounded. If the lattice, probe and launch
circuits each cost \(O(\log(2N_{\mathcal E}))\) gates up to precision
logarithms, the quantum setup therefore depends only logarithmically
on the exterior size. The subsequent classical march costs
\(\widetilde O(N_T)\) for a fixed core and interface. Further
trajectories on the same cut and harmonic exterior can reuse the
material response, with a separate incident force for each new launch.

\paragraph{Gain from sharing interval measurements}
The last two rows of Table~\ref{tab:simplified} give two ways to
supply the same response at the Verlet time steps. Chebyshev
interpolation measures the response at \(M+1\) nodes and yields a
polynomial that can be evaluated at any lag in \([0,T]\). If only
the Verlet times are needed, we can exploit the fact that the
normalized response at lag \(t_n\) is an average over \([0,t_n]\).
Section~\ref{sec:dyadic} reuses the same interval means across many lags and sets their precision according to their weights. Together these steps remove the factor \(M+1\) from the quantum query
bound. At fixed time step, interface, frequency scales and normalized
lag tolerance, extending the horizon \(T=N_T\Delta t\) gives
\(\widetilde O(N_T^{3/2})\) queries for Chebyshev interpolation and
\(\widetilde O(N_T)\) for shared intervals. 

For a specified microscopic thermal launch, the initial velocity norm \(\|\dot{\bm y}(0)\|\) can grow as \(\sqrt{N_{\mathcal E}}\), increasing the gain in \eqref{eq:total-gain}. The input model also requires an efficient preparation of the launch.  When the output is an ensemble statistic, a
classical phonon bath can instead supply boundary forces with the
prescribed statistics~\cite{KarpovParkLiu2007,LiE2007ThermalBath}.

\section{Numerical results}\label{sec:numerics}

In this section, we present three numerical experiments to test how accurately the two coupling algorithms reproduce defect motion and to estimate the cost of obtaining the required interface forces. We compute all trajectories classically and emulate the quantum measurements. The first experiment compares constructions of the exterior response for a wave packet scattered by a nonlinear bond. The next two examine how normalization affects concurrent traction readout and quantify the quantum setup and classical marching costs for a graphene trajectory at a prescribed accuracy. For comparisons at a relative tolerance, we divide the maximum norm of the core-displacement error by the maximum norm of the reference core displacement over the same time interval,
\begin{equation}
 E_q=\frac{\max_{0\le n\le N_T}\|\widehat{\bm q}_n-\bm q_n^{\rm ref}\|_2}
              {\max_{0\le n\le N_T}\|\bm q_n^{\rm ref}\|_2}.
 \label{eq:core-error}
\end{equation}

\subsection{Wave scattering by a local defect}\label{sec:incoming-test}

We first consider a periodic \(256\times192\) square lattice of unit
masses, with axial and diagonal spring stiffnesses \(1\) and \(1/4\).
The two atoms at \((0,0)\) and \((1,0)\) form the nonlinear core
\(\mathcal C\). Their bond has an additional quartic energy, and the
remaining atoms form the harmonic exterior \(\mathcal E\). A wave
packet centered at \(x=39\), uniform in the transverse direction,
travels toward the defect in the negative \(x\)-direction. Removing
the quartic term changes the core trajectory by \(16.0\%\).

We first test whether the discrete DtN map captures a wave arriving
from the exterior. At \(T=36\) and \(\Delta t=0.15\), the matched-Verlet
response in \eqref{eq:shifted-both}, with degree \(K=64\) and
deterministically evaluated moments, reproduces the full-scale Verlet
core trajectory to \(E_q=4.14\times10^{-11}\). Omitting the incident
field gives \(E_q\simeq1\). The same distinction appears in a direct
Verlet calculation: clamping the lattice at 32 layers excludes part of
the incident packet and gives \(E_q=1.68\), whereas 64 layers recover
the reference trajectory to roundoff. Both cuts hold omitted atoms at
their launch displacements. If the known incident wave is subtracted
first, eight buffer layers suffice to propagate the scattered field
with \(E_q=5.19\times10^{-7}\).

\begin{figure}[!htb]\centering
\includegraphics[width=.96\linewidth]{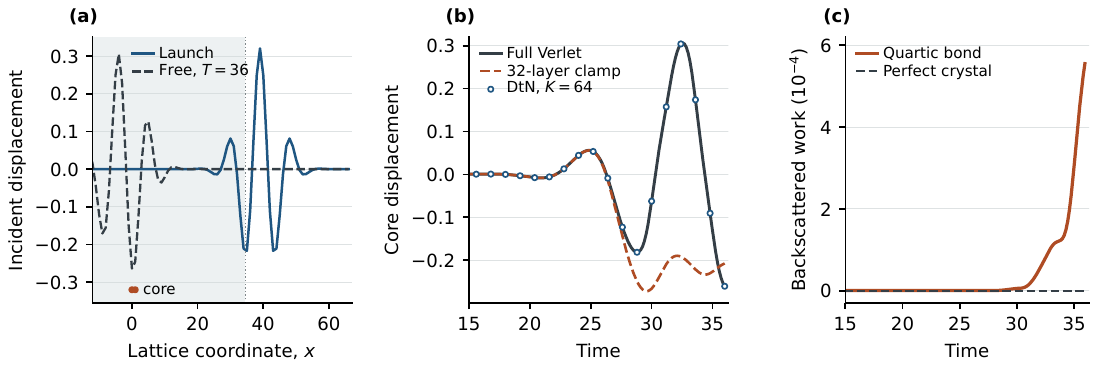}
\caption{Scattering by the quartic bond. (a) Incident packet at launch
and at \(T=36\). Shading marks the 32-layer domain. (b) Displacement
of the left core atom from full-scale Verlet, clamped Verlet and the
discrete DtN map with incident data. (c) Cumulative work carried back
across \(x=4.5\) by the scattered field.}
\label{fig:packet}
\end{figure}

The scattered field carries energy back toward the launch side
(Fig.~\ref{fig:packet}c). To isolate the effect of a returning wave on
the defect, we place reflecting sidewalls at distances of 4.5 or 32.5
lattice spacings from the bond. The incident motion at the core is the
same in both strips. With the quartic bond present, the core
trajectories differ by \(2.52\times10^{-3}\) in peak-relative norm
through \(T=48\). This difference measures the back action of the
reflected scattered field.

\paragraph{Quantum work at matched core-displacement accuracy}
We next compare three ways to construct the response table for the
same packet. After subtracting the incident wave, symmetry confines
the scattered core motion to the direction
\(\bm e=(1,-1)^\top/\sqrt{2}\), so one probe pair suffices.
For the continuous response we use \(C\bm e/\|C\bm e\|\) in
\eqref{eq:lag-probe} and estimate its lags by Chebyshev interpolation
or shared interval averages. For matched Verlet we estimate the
stiffness moments \(\bm p^\dagger T_k(2A/\alpha^2-I)\bm p\) of the
same probe \(\bm p\) and combine them to obtain the discrete lags
in \eqref{eq:shifted-both} (Appendix~\ref{app:stiffness-moments}).

\begin{figure}[!htb]\centering
\includegraphics[width=.98\linewidth]{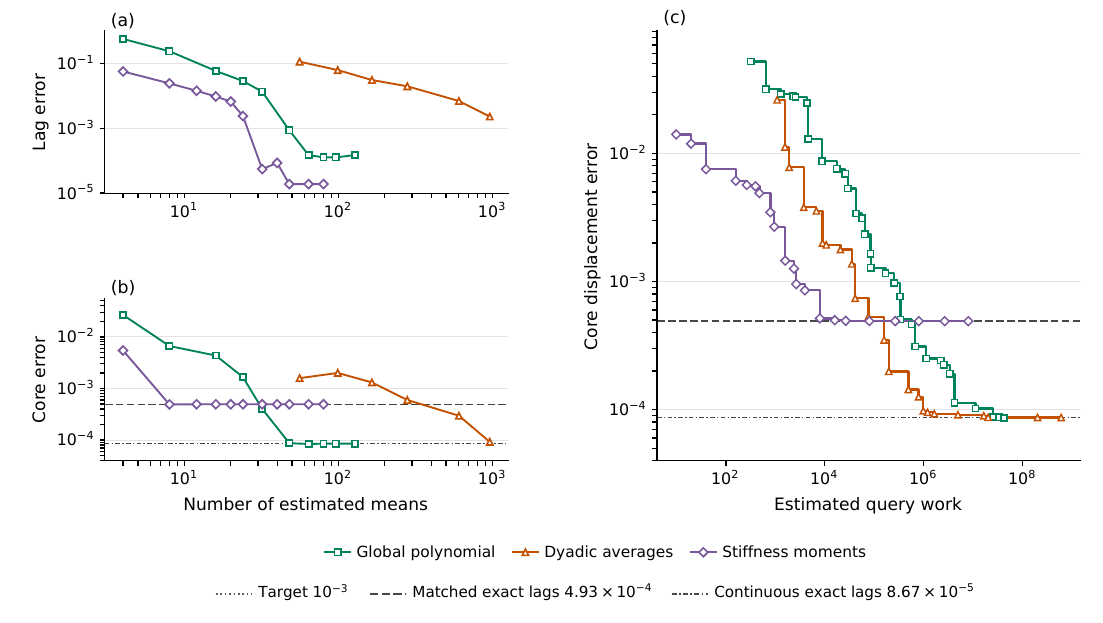}
\caption{Response estimates for the scattered packet over 960 steps.
(a) Maximum normalized lag error. (b) Maximum two-atom displacement
error against full-scale Verlet at \(\Delta t/16\). (c) Lowest observed
displacement error within each estimated query budget as resolution
and measurement precision vary. Each schedule is scored by its largest
error over three bounded deterministic roundings. Horizontal lines in
(b,c) mark the target and the errors obtained with exact lags.}
\label{fig:continuous-scattering}
\end{figure}

Figure~\ref{fig:continuous-scattering} compares the three methods at
\(T=36\) and \(\Delta t=0.0375\). The work estimate weights each
measured mean by its circuit depth and inverse absolute precision.
At an absolute core-displacement tolerance of \(10^{-3}\), eight
stiffness moments give the lowest estimated work among the tested
schedules: \(2.67\times10^3\), compared with \(4.13\times10^4\) for
shared intervals and \(2.54\times10^5\) for Chebyshev-time interpolation.
The latter supplies all 960 lags from 48 measured times and the known
zero-lag value. At the tighter tolerance of \(10^{-4}\), shared
intervals give the lowest estimated work, \(9.86\times10^5\).
The exact-lag calculations explain the change: at this step size,
matched Verlet already has a displacement error of \(4.93\times10^{-4}\)
against the fine-step reference, while the continuous response with
trapezoidal history gives \(8.67\times10^{-5}\).

\subsection{Concurrent force-readout diagnostic}\label{sec:gtest}

We consider a periodic graphene sheet with 144 atoms. The nonlinear core
\(\mathcal C\) contains 16 atoms surrounding a vertical defect bond
\((i,j)\), and the remaining atoms form the harmonic exterior
\(\mathcal E\). The potential consists of the Tersoff-derived harmonic
stiffness \(\mathsf K\), with three first and six second neighbors per
atom~\cite{tersoff1988,li2025}, and a quartic energy in the extension
of the defect bond,
\begin{equation}
 V=\tfrac12\bm d^\top\mathsf K\bm d
   +\tfrac{\beta}{4}\left(
      \|\bm R_j-\bm R_i+\bm d_j-\bm d_i\|
      -\|\bm R_j-\bm R_i\|\right)^4 .
 \label{eq:benchmark-energy}
\end{equation}
Here \(\bm d\) is the physical displacement and \(\beta=1500\) in
stiffness units per squared angstrom. Since the quartic bond lies
inside \(\mathcal C\), the exterior is harmonic by construction. We displace the bond ends by
\(\pm(0.015,0.075)\)~\AA{}, set all other displacements and all
velocities to zero, and take 120 Verlet steps of length \(0.01\) in
the reduced time units set by \(\mathsf K\) and unit mass. Removing
the quartic force changes the core trajectory by \(34.1\%\) in the
peak-relative measure~\eqref{eq:core-error}.

\begin{figure}[!htb]\centering
\includegraphics[width=.94\linewidth]{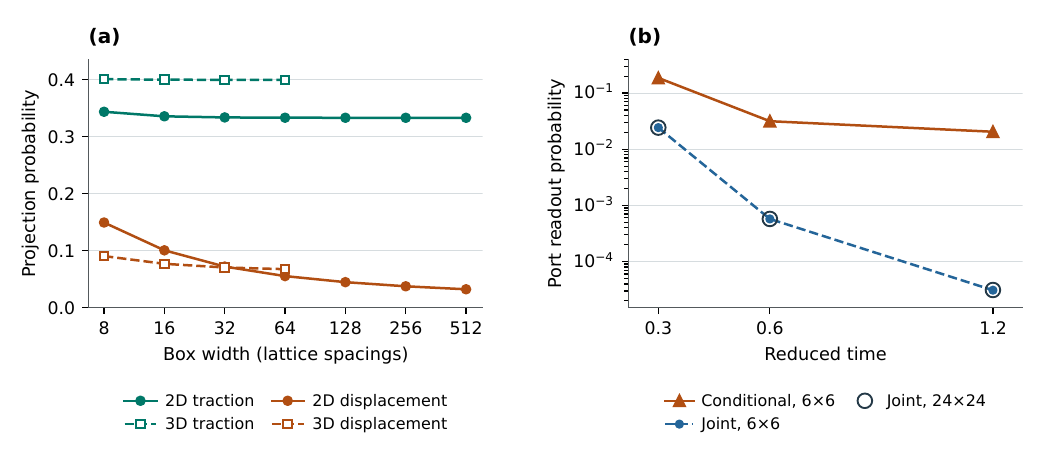}
\caption{Force-readout probabilities. (a) Conditional projection
probabilities for strain and displacement probes of an equilibrium
dipole in two- and three-dimensional spring lattices. (b) Displacement
readout of the vertical force contribution \((C^\top\bm y_m)_a\) at
the upper graphene defect atom. Triangles give the conditional
probability for 144 atoms. Filled and open circles give the joint
probability for 144 and 2,304 atoms, respectively.}
\label{fig:norm}
\end{figure}

We study how the success probability of the force readout depends on
the probe, the size of the exterior and the elapsed time. For the
strain readout, Eq.~\eqref{eq:strain-projection} factors the joint
probability into the probability of preparing the physical strain
and the conditional probability of projecting it onto the force
probe. The displacement readout has the same two factors
(Appendix~\ref{sec:direct}). For an equilibrium dipole with a fixed
interface, the two-dimensional strain probe retains a conditional
probability approaching \(1/3\) as the lattice grows, while the
displacement-probe probability decreases (Fig.~\ref{fig:norm}a).

Along the graphene reference trajectory, \((C^\top\bm y_m)_a\) is
close to \(0.063\) at both \(t=0.30\) and \(t=1.20\), although it
changes sign in between. Over this interval both probability factors
decrease, and their product falls by a factor of 787 in the 144-atom
sheet. Thus two nearly equal forces give signed overlaps differing
about 28-fold in magnitude, with a corresponding increase in coherent
estimation cost at fixed absolute force tolerance. Enlarging the sheet
to 2,304 atoms has little effect on the joint probability at the three
sampled times (Fig.~\ref{fig:norm}b). These measurements use a
prescribed Verlet trajectory to isolate the readout in
Algorithm~\ref{alg:march}.

\subsection{Quantum and classical overhead for a matched graphene trajectory}
\label{sec:matched-work}

We increase the periodic graphene sheet from 144 to 9,216 atoms while
keeping the 16-atom core and bond launch fixed. At each size, we compare
the reduced trajectory with full-scale Verlet using the same fitted
exterior stiffness, tabulated core and crossing blocks, and 120 steps
of length \(\Delta t=0.01\). All 60 simulated moment tables at each
size met the target \(E_q\le10^{-3}\) for the 32 core coordinates.
The largest observed error was \(1.91\times10^{-4}\).

\begin{table}[!htb]\centering\small
\caption{Marching work in millions of multiply-accumulates (MACs) and
stored coefficients for the same 16-atom core. Work counts stiffness
and response products. Full-scale storage counts the assembled stiffness
matrix; a periodic implementation can store the repeating stencil once.}
\label{tab:graphmatched}
\begin{tabular}{@{}r rr rr@{}}\toprule
 & \multicolumn{2}{c}{March (\(10^6\) MACs)}
 & \multicolumn{2}{c}{Stored coefficients}\\
\cmidrule(lr){2-3}\cmidrule(l){4-5}
Sheet atoms & Full-scale & Reduced & Full-scale & Reduced\\\midrule
   144 &   4.361 & 1.896 &    36,040 & 15,800\\
   576 &  20.252 & 1.896 &   167,368 & 15,800\\
 2,304 &  83.814 & 1.896 &   692,680 & 15,800\\
 9,216 & 338.065 & 1.896 & 2,793,928 & 15,800\\\bottomrule
\end{tabular}
\end{table}

The reduced march depends only on the 32 interface coordinates, so
its work stays at 1.896 million MACs. On the largest sheet, full-scale
Verlet requires 178 times as many. Constructing the response table
costs \(4.58\times10^{11}\) ideal controlled walks at each size:
the estimator resolves 528 interface pairs at 14 moment orders using
15 phase qubits and 63 repetitions per entry. Subsequent trajectories
on the same cut and exterior can reuse this table.

The factor of 178 measures the saving in marching work relative to
the full sheet. Over this short horizon, a classical calculation can
also exploit locality: clamped Verlet on 118 atoms of the largest
sheet meets the same target in 1.925 million MACs
(\(E_q=5.85\times10^{-4}\)).

\section{Discussion and outlook}\label{sec:conclusion}

A persistent computational challenge in the mechanics of materials is to capture how a lattice defect scatters phonons into the surrounding crystal and how the returning waves alter its motion. We have developed two quantum--classical frameworks for coupling the nonlinear defect core to this harmonic exterior. In concurrent coupling, the exterior returns the \(N_I\) components of the traction generated by the actual interface motion at each step. This can be economical for a single short trajectory when the history is inexpensive to replay, the preparation scale \(\|\bm\ell_a\|S_m\) permits accurate force readout, and a reusable table would require many distinct load--readout pairs. A dense reciprocal interface response, for example, contains \(N_I(N_I+1)/2\) such pairs. For longer trajectories, replaying the loading history at every exchange contributes a factor \(N_T^2\) to the concurrent query bound. Offline precomputation with shared intervals recovers the entire lag table in \(\widetilde O(N_T)\) queries at fixed time step, interface, encoding scale and normalized lag tolerance. Section~\ref{sec:fixedbudget} relates the force and lag tolerances to the same core-displacement accuracy.

The value of offline precomputation is especially clear when several defect trajectories share the same harmonic exterior. This reuse has a precedent in constitutive modeling. Garc\'ia-Cervera, Ren, Lu and E tabulated the Cauchy--Born energy of fcc aluminum and its strain derivatives for subsequent continuum calculations~\cite{garciacervera2008}. Guo, Roko\v{s} and Veroy built a reduced basis from microscopic stress fields for repeated constitutive evaluations~\cite{guo2021rb}. Bhattacharya et al. learned the dependence of homogenized stress on strain history and microstructure from cell problems~\cite{bhattacharya2026memory}. In our setting, the reusable quantity is the impulse response between interface atoms in \eqref{eq:ct-traction}. Linearity and time-translation invariance allow the same response to supply the traction for every displacement history on a fixed cut, with a separate incident force for each exterior launch. If a dislocation glides or a crack advances beyond the nonlinear region, the calculation must move the cut or update the reference crystal. How to transfer the existing wave field and its force history to the new partition remains an open question.

The trajectory generated by one microscopic thermal launch and the mean of a defect observable over an ensemble present different accuracy questions. A material table reused for many launches carries the same measurement error into every force history. Averaging over launches reduces sampling fluctuations, while any bias from that particular table persists. A weak-error analysis should distinguish these effects for observables such as bond extension or the populations of two defect configurations. The temperature dependence of their equilibrium population ratio gives access to an entropy difference, which can affect defect stability~\cite{KimTadmor2014,leimkuhler2016averages,braun2020rate}.

Quantum algorithms for linear dynamics already admit provable advantages for selected observables, including an exponential query separation for kinetic energy in certain harmonic networks~\cite{babbush2023}. Algorithms for more general linear differential equations similarly give access to a state encoding the solution~\cite{berry2017history}. Our extension uses this capability when the nonlinear vector fields  are confined to a few coordinates and the linear surroundings act back on them. Such models also exist in other applications. For example, Fiorentini et al. retained atomistic interactions at a protein binding site and coupled it to an elastic network~\cite{fiorentini2020lysozyme}. Linearizing that network about a reference conformation would make its response to local bond stretching and bending accessible to the same construction. In aeroacoustics, Lim et al. coupled nonlinear Euler flow near a standing shock to linearized wave propagation farther away~\cite{lim2023acoustic}. There the interface must transmit pressure and velocity, and a reusable response would require a stationary exterior with an efficiently encoded propagator. In either application the relevant comparison is the cost of returning those interface data at the accuracy required by the nonlinear motion.
\section*{Acknowledgments}
The work is supported by NSF Grant DMS-2411120.

\FloatBarrier
\appendix
\renewcommand{\thesection}{\Alph{section}}
\makeatletter
\@addtoreset{figure}{section}\@addtoreset{table}{section}
\def\appendixprefix@section{Appendix~}
\let\normalnumberformat\@seccntformat
\renewcommand{\@seccntformat}[1]{\csname appendixprefix@#1\endcsname\normalnumberformat{#1}}
\makeatother

\section{Construction of the traction estimators}\label{app:readouts}

\subsection{Local traction and exterior energy}\label{sec:readout-prob}
For the compatible factors in~\eqref{eq:compatible-gram}, set
\(E_{\rm str,m}=\|\bm w_m\|^2/2\) and
\(\tau^{\rm str}_{m,a}=\bm\ell_a^\top\bm w_m\), with \(\bm\ell_a=L\bm e_a\).
For nonzero strain and probe, the squared overlap in~\eqref{eq:live-force} factors as
\begin{equation}
 p^{\rm strain}_{m,a}
  =\underbrace{\frac{2E_{\rm str,m}}{S_m^2}}_{\text{strain preparation}}\,
   \underbrace{\frac{|\tau^{\rm str}_{m,a}|^2}
       {2E_{\rm str,m}\|\bm\ell_a\|^2}}_{\text{projection given preparation}}
  =\frac{|\tau^{\rm str}_{m,a}|^2}{\|\bm\ell_a\|^2 S_m^2}.
 \label{eq:strain-projection}
\end{equation}
The second factor measures the fraction of strain energy in the probe direction. For a probe aligned with a single crossing spring of stiffness \(k\) carrying force \(f\), that fraction is \([f^2/(2k)]/E_{\rm str,m}\). For an algebraically decaying strain of order \(r^{-s}\), the energy beyond a cut of radius \(R_{\mathcal C}\) stays bounded as the outer radius grows if the shell energy is \(O(r^{d-1-2s})\) and \(2s>d\). A thermal launch, by comparison, can carry energy proportional to exterior volume.

For direct preparations, a bound \(V_*\) on the accepted half-step core speeds gives
\[
 S_m\le \|\bm w_0\|+
 b_{\Delta t}\bigl(\|\dot{\bm y}(0)\|+t_m\|LR\|V_*\bigr)
\]
by~\eqref{eq:strain-scale}. If this upper bound is independent of exterior size and the directional strain \(|\tau^{\rm str}_{m,a}|/\|\bm\ell_a\|\) stays bounded below at the steps of interest, then \(p^{\rm strain}_{m,a}\ge p_0>0\) there uniformly in exterior size. Equation~\eqref{eq:live-force} and coherent mean estimation give \(O(1+1/(r\sqrt{p_0}))\) preparations for relative tolerance \(r\) in the strain contribution at fixed confidence. For relative accuracy in the total traction, cancellation with the known correction in~\eqref{eq:live-force} requires an absolute force tolerance \(r|(\bm\tau_m)_a|\).

\subsection{Displacement readout for an independent crossing block}\label{sec:direct}
For an independently supplied \(C\), set \(\bm c_a=C\bm e_a\). The traction component is \((D_\times\bm u_{\Gamma,m})_a+\bm c_a^\top\bm y_m\). For \(m\ge1\) the solution of~\eqref{eq:verletbath} is
\[
 \bm y_m=T_m(X)\bm y_0+\Delta t\,U_{m-1}(X)\dot{\bm y}(0)
   -(\Delta t)^2\sum_{j=0}^{m-1}\nu_j U_{m-1-j}(X)C\bm u_{\Gamma,j},
 \qquad \nu_0=\tfrac12,\quad \nu_j=1\quad(j>0).
\]
The polynomials \(T_m(X)\) and \(U_{\ell-1}(X)/\ell\) are contractions on the stable step interval. With unit preparations of the nonzero sources, their linear combination gives the displacement signal \(\bm y_m/\Lambda_m\), where
\[
 \Lambda_m=\|\bm y_0\|+t_m\|\dot{\bm y}(0)\|
   +(\Delta t)^2\sum_{j<m}\nu_j(m-j)\|C\bm u_{\Gamma,j}\|.
\]
Interference with the unit probe \(\bm c_a/\|\bm c_a\|\) returns the signed exterior contribution divided by \(\|\bm c_a\|\Lambda_m\). For a nonzero field and probe the corresponding joint projection probability is
\[
 \frac{\|\bm y_m\|^2}{\Lambda_m^2}
 \frac{|\bm c_a^\top\bm y_m|^2}{\|\bm c_a\|^2\|\bm y_m\|^2}.
\]
This is the displacement diagnostic in Fig.~\ref{fig:norm}. The displacement norm can grow with specimen size even when the strain energy stays bounded.

\subsection{The driven strain and its preparation scale}\label{app:strain-increments}
The identities \(BX=X'B\) and \(C=B^\top L\) give the first Verlet step
\(\bm w_1=X'\bm w_0+\Delta t B\dot{\bm y}(0)+L\Delta\bm u_{\Gamma,0}\).
For \(n\ge1\), the recurrence from~\eqref{eq:verletbath} becomes
\[
 \bm w_{n+1}=2X'\bm w_n-\bm w_{n-1}
       +L(\Delta\bm u_{\Gamma,n}-\Delta\bm u_{\Gamma,n-1}).
\]
Each increment acts at successive steps with opposite signs, yielding \(U_{m-1-j}-U_{m-2-j}\) in~\eqref{eq:live-strain}. For the newest increment, \(U_0-U_{-1}=1\).

For \(z\in[-1,1]\), let \(c=\Delta t\,\alpha/2<1\), \(\theta=2\arcsin(cz)\), and \(x=\cos\theta\). The identities
\[
 (U_k-U_{k-1})(x)=\frac{\cos((k+1/2)\theta)}{\cos(\theta/2)},\qquad
 2czU_{m-1}(x)=\frac{\sin(m\theta)}{\cos(\theta/2)}
\]
bound both polynomials by \(b_{\Delta t}=(1-c^2)^{-1/2}\). The first is even in \(z\) and acts on the strain sector. The second is odd and its site-to-strain block is \(\Delta t\,U_{m-1}(X')B\). The initial-strain polynomial \(T_m(x)\) has norm at most one. After division by these bounds, each polynomial satisfies the parity and boundedness conditions for quantum singular value transformation and has degree at most \(2m\)~\cite{gilyen2019}.

The LCU index amplitudes are proportional to the square roots of the source norms times their respective polynomial bounds. A controlled choice of the source and its polynomial, followed by inversion of the index preparation, gives \(\bm w_m/S_m\) on the clean flags. Thus \(S_m\) in~\eqref{eq:strain-scale} is the coefficient one-norm. A subnormalized source block contributes its preparation scale. A controlled choice of polynomial sequences of degree at most \(2m\) uses \(O(m)\) generator queries. The local source preparations, phase controls and classical loading are charged separately.

The probe in~\eqref{eq:live-force} has clean flags, making it orthogonal to the failure component of the signal. With phase length \(M\), coherent amplitude estimation gives error at most \(2\pi/M+2\pi^2/M^2\) in the signed ancilla mean with probability at least \(8/\pi^2\)~\cite{brassard2002}. A block error \(\xi\) adds at most \(\xi\) to the mean. Multiplication by \(\|\bm\ell_a\|S_m\) converts both errors to force units.

\section{Trajectory error and computational work}\label{app:error-cost}

\subsection{The displacement gain}
We compare trajectories of the same harmonic-exterior model and with the same launch. We extend the computed core positions to the full crystal by advancing an exact Verlet exterior on the computed trace. Let \(\bm e_n\) be the difference between these full positions and those from full-scale Verlet. The core receives the traction discrepancy multiplied by \(-R^\top\), so its contribution to the acceleration error has norm at most \(r_{\rm num}\). With \(X_{\rm lin}=I-(\Delta t)^2\mathsf K_{\rm lin}/2\), subtraction of the position recurrences gives
\[
 \bm e_n=(\Delta t)^2\sum_{j=0}^{n-1}\nu_j
 U_{n-j-1}(X_{\rm lin})\,\delta\bm a_j,\qquad
 \|\delta\bm a_j\|\le L_{\rm nl}\|\bm e_j\|+r_{\rm num}.
\]
Here \(\delta\bm a_j\) includes the nonlinear force difference and the traction discrepancy. The first half kick gives \(\nu_0=1/2\), and \(\nu_j=1\) for \(j>0\). The stability hypothesis gives \(\|U_{n-j-1}(X_{\rm lin})\|\le n-j\). Replacing the propagator by this bound and taking equality in the scalar recurrence yields a nonnegative majorant for \(\|\bm e_n\|\). For \(L_{\rm nl}>0\) it is
\[
 \frac{r_{\rm num}}{L_{\rm nl}}
 \big[T_n(1+(\Delta t)^2L_{\rm nl}/2)-1\big].
\]
Since \(\operatorname{arcosh}(1+z^2/2)\le z\) for \(z\ge0\), the majorant is bounded by \(S_xr_{\rm num}\) on \([0,T]\). For \(L_{\rm nl}=0\) it is \(r_{\rm num}t_n^2/2\), giving \(S_x=T^2/2\). Projection onto the core proves \eqref{eq:traj} for the full-scale Verlet reference. Comparison with the continuous trajectory adds the full-scale Verlet discretization error \(E_{\Delta t}\). For concurrent readout, component errors at most \(\varepsilon_\tau\) give \(r_{\rm num}\le\sqrt{N_I}\,\varepsilon_\tau\). If the Lipschitz constant is local, these bounds apply for as long as both trajectories stay in its tube.

The error recurrence also gives an exact work balance. With \(\bm v^e_{n+1/2}=(\bm e_{n+1}-\bm e_n)/\Delta t\), set
\[
 E^e_{n+1/2}=\tfrac12\|\bm v^e_{n+1/2}\|^2
             +\tfrac12\bm e_{n+1}^\top\mathsf K_{\rm lin}\bm e_n .
\]
Taking the inner product of the recurrence with \((\bm e_{n+1}-\bm e_{n-1})/2\) after dividing by \((\Delta t)^2\) gives
\[
 E^e_{n+1/2}-E^e_{n-1/2}
 =\Delta t\left\langle\tfrac12(\bm v^e_{n+1/2}+\bm v^e_{n-1/2}),\delta\bm a_n\right\rangle .
\]
The energy is nonnegative under the stability hypothesis, since its two quadratic forms in midpoint variables have matrices \(\mathsf K_{\rm lin}\) and \(I-(\Delta t)^2\mathsf K_{\rm lin}/4\).

\subsection{Proof of Theorem~\ref{thm:cost}}\label{app:cost-proof}
\begin{proof}
On the simultaneous estimation event, the physical error in entry \((a,b)\) of \(C^\top F_\ell C\) is bounded by \(\eta\|C\bm e_a\|\,\|C\bm e_b\|\). The matrix of entrywise bounds is an outer product, giving operator-norm error at most \(\eta\|C\|_F^2\). Likewise, for an incident input \(\bm z\) the error in \(C^\top F_\ell\bm z\) is at most \(\eta\|C\|_F\|\bm z\|\). In \eqref{eq:shifted-both}, \(u_{\max}\) bounds the norm of the accepted trace, the velocity term is weighted by \(t_n\), and the history weights sum to \(\Delta t\sum_{j<n}\nu_j t_{n-j}=t_n^2/2\). Equation~\eqref{eq:traj} therefore bounds the displacement error due to the estimated lags by \(B_T^{\rm tot}\eta\), for any correlations among their errors. The static offset \(C^\top\bm y_0\) is supplied to its allocated accuracy. If it is measured separately, its preparation and measurement are charged with the launch data.

For matched Verlet, exact lags reproduce the full-scale Verlet force on the accepted trace. For continuous lags evaluated with the trapezoidal history in \eqref{eq:shifted-both}, we can also bound the difference from this discrete force. On a mode with \(x=\omega\Delta t\in[0,2]\), set \(u=x/2\), \(\theta=2\arcsin u\), \(s=\sqrt{1-u^2}\), and \(p_\ell=U_{\ell-1}(\cos\theta)/\ell\). Then \(|p_\ell|\le1\) and \(sp_\ell=\sin(\ell\theta)/(\ell x)\), with limiting values at zero. Consequently
\[
 |p_\ell-\operatorname{sinc}(\ell x)|
 \le (1-s)+(\theta/x-1)
 \le u^2+(\pi/2-1)u^2=(\pi/8)x^2 .
\]
The last inequality follows from the nonnegative power series of \(\arcsin(u)/u\). By the spectral theorem this bounds the error of the lag operator at every step by \((\pi/8)(\Omega\Delta t)^2\). Both force histories have the same weights, so the resulting displacement error is at most \(B_T^{\rm tot}(\pi/8)(\Omega\Delta t)^2\). This gives a sufficient allowance for the continuous-history discretization in addition to \(E_{\Delta t}\) when the reference is continuous.

The hypothesis leaves at least \(\varepsilon/2\) for lag estimation after accounting for time discretization and the supplied inputs. For \(B_T^{\rm tot}>0\), we take \(\eta=\min\{1/2,\varepsilon/(2B_T^{\rm tot})\}\). A zero gain requires no lag measurements. Proposition~\ref{prop:dyadic} gives the setup count in \eqref{eq:main-cost} and simultaneous success probability at least \(1-\delta\). Each prefix is assembled from one previous prefix and one stored block, giving \(O(N_{\rm pair}N_T)\) classical operations, also for the two parity sums of matched Verlet.

During the march we split the time interval recursively. Once the left half of an interval is known, a zero-padded FFT convolution adds its displacements to the forces in the right half. Each pair of distinct source and receiving steps enters at exactly one split. The intervals at each level are disjoint, and a block of length \(s\) costs \(O(N_I^2 s\log(2s))\) operations~\cite{hairer1985volterra}. Summing over levels and adding \(O(N_{\mathcal C})\) work per step for bounded-range core interactions gives the last line of \eqref{eq:main-cost}.
\end{proof}

\subsection{Finite-time truncation of the exterior in a classical Verlet calculation}\label{app:incoming-box}
Table~\ref{tab:simplified} counts the work of advancing every exterior atom. To compare the methods at the same accuracy in core displacement, we derive a sufficient number of exterior layers for a classical Verlet calculation through time \(T\). The material response and the incident force in \eqref{eq:shifted-both} involve different propagation paths. A wave emitted by the core must reach the artificial boundary and return before the core feels the truncation. A wave present in the initial data can travel directly to the core.

A practical case for response precomputation arises when a wave reflected by a distant physical boundary affects the defect within the observation time, as in the scattering test of Section~\ref{sec:incoming-test}. A direct Verlet calculation propagates the scattered wave anew for each core trajectory. For a fixed exterior and cut, Algorithm~\ref{alg:reusable} includes these reflections in the interface response \(C^\top P_\ell C\) measured once. Runs with different initial defect velocities can reuse this response and, for the same exterior initial data, the same incident force.

For matched Verlet, let the retained exterior sites be those within \(R_{\rm box}\) edges of the support of \(C\) in the stiffness graph. Holding all other sites at their launch displacements makes the increment \(\bm y_n-\bm y_0\) vanish there. The retained increment evolves under the principal submatrix of \(A\), with the restricted initial velocity and the restriction of the full launch load \(-A\bm y_0\). Evaluating this load needs one additional layer of initial data. The static offset \(C^\top\bm y_0\) is common to the two boxes.

The generating function \(((1-z)^2I+z(\Delta t)^2 A)^{-1}\) for the Verlet impulse gives
\[
 P_\ell=\frac1\ell\sum_{k=0}^{\ell-1}(- (\Delta t)^2A)^k\binom{\ell+k}{2k+1}.
\]
Since \(\binom{\ell+k}{2k+1}/\ell=\prod_{j=1}^k(\ell^2-j^2)/(2k+1)!\), the uniform operator error after degree \(K\) is at most \(\sum_{k>K}(\alpha T)^{2k}/(2k+1)!\le4^{-(K+1)}e^{2\alpha T}\), also for the principal submatrix. For a prescribed tail bound \(\eta_{\rm I}\), a degree of order \(1+\alpha T+\log(1/\eta_{\rm I})\) suffices, capped at \(N_T-1\) where the expansion is exact.

A graph path from the support of \(C\) to an omitted site and back requires at least \(2(R_{\rm box}+1)\) edges. Thus \(R_{\rm box}=\lfloor K/2\rfloor\) preserves all degree-\(K\) matrix elements between interface loads. This radius suffices for the full traction estimate when \(\bm y_0=\dot{\bm y}(0)=0\). For general exterior initial data, a path can start outside the retained layers, and \(R_{\rm box}=K\) suffices. The difference between exact interface entries then comes only from the two truncation tails. Comparing the two exterior forces on the same trace in \eqref{eq:shifted-both} gives
\[
 r_{\rm box}\le2\eta_{\rm I}\|C\|_2\left[T\|\dot{\bm y}(0)\|
 +\tfrac12T^2\big(\|A\bm y_0\|+\|C\|_2u_{\max}\big)\right].
\]
Equation~\eqref{eq:traj} bounds the resulting discrepancy against full-scale Verlet by \(S_xr_{\rm box}\). In a fixed-dimensional crystal with bounded interaction range, a fixed interface has \(O((1+K)^d)\) sites within \(K\) layers. At fixed \(T\), bounded physical gains and locally accessible launch data, choosing \(\eta_{\rm I}\) proportional to \(\varepsilon\) and \((\Delta t)^2\asymp\varepsilon\) therefore gives \(\widetilde O(\varepsilon^{-1/2})\) classical work.

\section{Estimation of the reusable response}\label{app:reusable-estimation}
\subsection{Circuit and interpolation proof for the Chebyshev-time moments}\label{app:time-estimator}
\begin{proof}[Proof of Theorem~\ref{thm:time-moments}]
The factor encoding and its inverse, controlled on the site and strain sectors of \eqref{eq:generator}, give an encoding of \(H/\alpha\) after a sector swap. At a node \(s>0\) we prepare a uniform superposition over \(N_{\rm quad}\) midpoint times \((k+1/2)s/N_{\rm quad}\), with \(N_{\rm quad}\) a power of two. Each address controls the corresponding evolution. Unpreparing the address register gives their average in its clean block. Between site probes the odd powers of \(H\) vanish, so this implements midpoint quadrature for \eqref{eq:lag-probe}. The bound \(\|\partial_u\cos(uH)\|_2\le\Omega\) gives quadrature error at most \(\Omega s/(2N_{\rm quad})\). With \(N_{\rm quad}\ge\max\{1,2\Omega s/\sigma\}\) this is at most \(\sigma/4\). The binary-controlled durations, including the midpoint offset, sum to less than \(s\). Qubitization thus uses \(\widetilde O(1+\alpha s)\) generator queries to implement the average to an additional error of at most \(\sigma/4\)~\cite{low2019}.

In a Hadamard test the reference branch prepares \(|0,\bm p\rangle\), with all flags clean, and the signal branch applies the averaging circuit to \(|0,\bm z\rangle\). If \(V\) is the circuit's clean time-register block, the probability of the \(+\) outcome is \((1+\operatorname{Re}\langle0,\bm p|V|0,\bm z\rangle)/2\). Coherent amplitude estimation resolves the signed overlap to the remaining error allowance in \(\widetilde O(1/\sigma)\) calls to the preparation and its inverse~\cite{brassard2002}. A reference phase gives the imaginary part when needed. If a nonzero probe column \(C\bm e_a\) of known norm is supplied by an encoding of \(C/\beta_C\), each normalized preparation requires \(O(\beta_C/\|C\bm e_a\|)\) uses of the encoding and its inverse by amplitude amplification. At \(s=0\) the circuit estimates only the probe overlap. Failure probability \(\delta/[N_{\rm pair}(M+1)]\) per pair and node suffices for joint confidence \(1-\delta\). Since \(\sum_j s_j=(M+1)T/2\), the total is \(\widetilde O(N_{\rm pair}(M+1)(1+\alpha T)/\sigma)\) generator queries and \(\widetilde O(N_{\rm pair}(M+1)/\sigma)\) probe preparations and inverses.

For an eigenvalue \(|\lambda|\le\Omega\), set \(v(t)=\operatorname{sinc}(\lambda t)=\int_0^1\cos(r\lambda t)\,dr\). The zero mode has \(v\equiv1\). Let \(I_M\) denote exact interpolation at the stated nodes. We give three bounds for \(\|v-I_Mv\|_{\infty,[0,T]}\).

First write \(t=T(1-\cos\theta)/2\) and \(g(\theta)=v(t)\). The function \(g\) is even and \(2\pi\)-periodic. The integral gives \(|v'|\le|\lambda|/2\), and \((tv)''=-\lambda\sin(\lambda t)\) gives \(|tv''|\le2|\lambda|\). Since \((t')^2=t(T-t)\) and \(|t''|\le T/2\), the chain rule yields \(\|g''\|_\infty\le 9|\lambda|T/4\). Favard's periodic approximation theorem gives a trigonometric polynomial of degree at most \(M\) with uniform error at most \(\pi^2\|g''\|_\infty/[8(M+1)^2]\)~\cite{favard1937}. Averaging the approximant at \(\theta\) and \(-\theta\) preserves this bound and leaves a cosine polynomial, hence a polynomial in \(1-2t/T\). The norm of \(I_M\) is \(\Lambda_M\). Since it reproduces the approximating polynomial, the interpolation error is bounded by \(1+\Lambda_M\) times its approximation error. This gives the first term in the minimum.

For the second bound, the Bernstein ellipse with parameter two in \(x=2t/T-1\) has \(|\operatorname{Im}t|\le3T/8\). The integral for \(v\) bounds its modulus there by \(e^{3\Omega T/8}\). Summing the Chebyshev tail and its aliases at the Lobatto nodes gives \(4e^{3\Omega T/8}2^{-M}\)~\cite[Ch.~8]{trefethen2019atap}. For the third bound, differentiation under the integral gives \(\|v^{(k)}\|_\infty\le|\lambda|^k/(k+1)\) on the real axis. The monic nodal polynomial in \(x\) is \(2^{1-M}(x^2-1)U_{M-1}(x)\). Its modulus is at most \(2^{1-M}\), since \((1-x^2)U_{M-1}(x)=\sin\theta\sin(M\theta)\) at \(x=\cos\theta\). Rescaling gives \(\max_{[0,T]}|\prod_{j=0}^M(t-s_j)|\le4(T/4)^{M+1}\). The real scalar interpolation remainder of order \(M+1\) now gives \(4(\Omega T/4)^{M+1}/(M+2)!\).

The cardinal weights are scalar and common to all modes. The spectral theorem transfers the worst scalar bound to operator norm, and taking a matrix element between unit site probes preserves it for real or complex probes. Nodal errors add at most \(\Lambda_M\sigma\), proving \eqref{eq:global-error}. For Lobatto nodes \(\Lambda_M=O(1+\log(M+1))\)~\cite[Ch.~15]{trefethen2019atap}. With \(x=\Omega T/\eta\), the first bias bound admits a sufficient degree \(O(1+\sqrt{x\log(2+x)})\), and the ellipse bound admits \(O(1+\Omega T+\log(1/\eta))\). Taking the smaller sufficient degree bounds the chosen \(M\) from above. Substitution of \(\sigma=\eta/(2\Lambda_M)\) in the circuit counts proves \eqref{eq:time-sample-cost}.

To see why the square root occurs, take \(B=(\Omega)\) with identical unit site probes. Then \(f(t)=\operatorname{sinc}(\Omega t)\). For \(\Omega T\ge\pi\), \(f\) falls from one to zero between \(0\) and \(\pi/\Omega\). A real polynomial \(P\) of degree \(M\) with uniform error \(\eta<1/2\) must satisfy \(\|P'\|_\infty\ge(1-2\eta)\Omega/\pi\). Markov's inequality on \([0,T]\) gives \(\|P'\|_\infty\le 2M^2(1+\eta)/T\). Thus \(M^2\ge (1-2\eta)\Omega T/[2\pi(1+\eta)]\). The real part gives the same conclusion for complex polynomials. This bounds the degree required on the full time interval.

For the variance bound, let \(\ell_j(t)\) be the cardinal weights of the global interpolant and set \(w_0=w_M=1/2\), \(w_j=1\) for \(0<j<M\). With \(t=T(1-\cos\theta)/2\), discrete cosine orthogonality at the Lobatto nodes gives
\[
 \sum_{j=0}^M\frac{\ell_j(t)^2}{w_j}
 =\frac{2}{M}\sum_{k=0}^M w_k\cos^2(k\theta)\le 2.
\]
Independent nodal estimates contribute \(\sum_j\ell_j(t)^2\operatorname{Var}[\widehat f(s_j)]\le 2v\) to the variance at \(t\). For unbiased estimates the remaining mean-square error is the squared exact-data interpolation error.
\end{proof}

\subsection{Sharing interval averages}\label{app:dyadic-proof}
\begin{proof}[Proof of Proposition~\ref{prop:dyadic}]
For a block \(\mathcal I=[i,j)\) of \(\kappa=j-i\) consecutive steps we average the site
block of the propagator over \([t_i,t_j]\). On a mode of frequency \(\omega\)
the resulting filter is
\[
   G_{\mathcal I}(\omega)=\frac{1}{\kappa\Delta t}\int_{t_i}^{t_j}\cos(s\omega)\,ds
   =\cos\!\big((t_i+t_j)\omega/2\big)
      \operatorname{sinc}\!\big(\kappa\Delta t\,\omega/2\big).
\]
It has magnitude at most one, including \(G_{\mathcal I}(0)=1\).
Each prefix \([0,n)\) is a disjoint union of at most
\(J=\lceil\log_2(2N_T)\rceil\) aligned blocks of power-of-two length.
Its normalized lag is the sum of their averages with weights \(\kappa/n\).
On a block ending at \(j\) we allow absolute error
\(\eta_j=\eta j/(J\kappa)\). When \(\eta_j\ge1\), zero already meets this
allowance. Since \(j\le n\) in every prefix using the block, its
contribution to that prefix's error is at most \(\eta/J\).

Quantum singular value transformation~\cite{gilyen2019} implements the filter as a bounded even polynomial of the encoded generator \(H/\alpha\).
Indeed, \(G_{\mathcal I}(\alpha x)\) is real and even on \([-1,1]\), and its integral
representation bounds its modulus at complex \(x\) by
\(\exp(\alpha t_j|\operatorname{Im}x|)\).
Chebyshev truncation on a fixed Bernstein ellipse therefore needs degree
\(O(1+\alpha t_j+\log(1/\eta_j))\) for error \(\eta_j/4\).
Scaling the polynomial by \(1/2\) makes it a bounded signal transformation.
The known factor of two is restored in signed overlap estimation.
Coherent mean estimation takes \(\widetilde O(1/\eta_j)\) uses of this
circuit and its inverse. At fixed length \(\kappa\), the right endpoints are
\(j=k\kappa\). Up to logarithms the query count for a probe pair on that
scale is bounded by
\[
   \sum_{k=1}^{\lfloor N_T/\kappa\rfloor}
      \frac{J(1+\alpha k\kappa\Delta t)}{k\eta}
   \le \frac{J}{\eta}\big(1+\log N_T+\alpha T\big).
\]
There are at most \(J\) scales. For the canonical block at endpoint \(j\), its width \(\kappa_j\) is the largest power of two dividing \(j\), so \(k=j/\kappa_j\) is odd. The sum over all positive \(k\) used above is an upper bound. Once the estimates have been recorded, \(h_j=h_{j-\kappa_j}+\kappa_j\widehat g_{j-\kappa_j,j}\) forms each unnormalized prefix in constant arithmetic work, with \(h_0=0\). Generating block widths and the full table takes \(O(N_T)\) operations per pair. A union bound over measured
blocks and probe pairs gives the simultaneous confidence.

For Verlet, write \(X=I-(\Delta t)^2 A/2\). The finite trigonometric sum
for \(U_{n-1}(\cos\theta)\) gives, including endpoint values by continuity,
\[
  P_{2m}=\frac{1}{m}\sum_{j=0}^{m-1}T_{2j+1}(X),\qquad
  P_{2m+1}=\frac{I+2\sum_{j=1}^{m}T_{2j}(X)}{2m+1}.
\]
The two parities require the same single pass on their respective prefix arrays. Each involves a prefix average. If the identity overlap of the two probes is unknown, one signed overlap estimate to accuracy \(\eta\) supplies it for every odd lag, at an additional \(\widetilde O(1/\eta)\) calls per pair within the stated bound. In the second formula the identity
overlap and the average have nonnegative weights summing to one.
For a block ending at index \(j\) the largest mode \(r\) is at most \(2j\).
On the encoded variable \(x\in[-1,1]\), this mode is
\(T_r(1-2c^2x^2)=\cos(2r\arcsin(cx))\).
Here \(c=\alpha\Delta t/2\le1\). If \(c\ge1/2\), the exact block polynomial has degree \(O(j)=O(jc)\).
If \(c<1/2\), the Bernstein ellipse of parameter two satisfies
\(|x|\le5/4\), so \(|cx|\le5/8\) and the convergent power series bounds
\(|\arcsin(cx)|\le {\rm const.}\,c\). Every mode in the block, and hence its
average, is bounded by \(\exp(O(jc))\) on this ellipse. The same truncation
and rescaling now need degree \(O(1+jc+\log(1/\eta_j))\).
Since \(jc=\alpha t_j/2\), the preceding sum applies to both parities.
\end{proof}

For a single normalized lag the factor \(\alpha T/\eta\) can already be
necessary under black-box access. Consider the scalar factor \(B=\omega\)
and unit probes. Near \(\omega T=\pi\), two frequencies separated by
\(c_0\eta/T\), for a suitable fixed \(c_0\), give values of
\(\operatorname{sinc}(\omega T)\) separated by more than \(2\eta\).
For \(\alpha T\ge2\pi\) and sufficiently small \(\eta\), one may encode
these factors by rotations with signal \(\omega/\alpha\) whose operator
distance is \(O(\eta/(\alpha T))\). Adding the site--strain swap encodes
the Hermitian generator at the same distance. Telescoping \(Q\) oracle
calls bounds the distance of the two output states by
\(O(Q\eta/(\alpha T))\). Thus constant distinguishing advantage requires at least a constant times \(\alpha T/\eta\) queries. This lower bound concerns an unknown
factor supplied through an oracle.

\subsection{Accuracy and implementation of stiffness moments}
\label{app:stiffness-moments}

Let \(I_K\) denote interpolation at the \(K+1\) stiffness nodes of
Section~\ref{sec:stiffness-moments}. In the variable \(y=2\lambda/\alpha^2-1\)
of~\eqref{eq:sm-moments}, the continuous operator \(F(t)\) in~\eqref{eq:lag-probe}
and the matched-Verlet operator \(P_\ell\) in~\eqref{eq:shifted-both} have
the scalar filters
\[
 g_t^{\mathrm{cont}}(y)=
   \operatorname{sinc}\!\left(\alpha t\sqrt{(1+y)/2}\right),\qquad
 g_\ell^{\mathrm V}(y)=\frac1\ell U_{\ell-1}\!\left(1-c^2(1+y)\right),
 \qquad c=\alpha\Delta t/2\leq1,\quad \ell\geq1 .
\]
Thus \(F(t)=g_t^{\mathrm{cont}}(Y)\) and \(P_\ell=g_\ell^{\mathrm V}(Y)\).

\paragraph{Effect of errors in the measured moments}
For either filter \(g\), write \(I_Kg=\sum_{k=0}^K a_kT_k\). Then
\begin{equation}
 (-1)^k a_k\geq0,\qquad \sum_{k=0}^K|a_k|=1 .
 \label{eq:sm-coeff-mass}
\end{equation}
To see this, put \(y=-\cos\theta\). For the matched lag, the unit vectors
\(\bm v_\theta=(c\cos\theta,c\sin\theta,\sqrt{1-c^2},0)\) on \(S^3\) give
\[
 g_\ell^{\mathrm V}(-\cos(\theta-\phi))
   =\ell^{-1}U_{\ell-1}(\bm v_\theta\cdot \bm v_\phi).
\]
The spherical-harmonic addition formula on \(S^3\) expresses this kernel
as a positive multiple of a sum of products
\(Y_{\ell-1,r}(\bm v_\theta)\overline{Y_{\ell-1,r}(\bm v_\phi)}\), since
\(C_m^{(1)}=U_m\). Hence its Gram matrices are positive semidefinite.%
\footnote{H.~S. Cohl and R.~M. Palmer, \emph{Fourier and Gegenbauer
expansions for a fundamental solution of Laplace's equation in
hyperspherical geometry}, SIGMA 11 (2015), 015, Eq.~(4.7) with \(d=4\).
DOI 10.3842/SIGMA.2015.015.}
For fixed \(t\), take \(\Delta t=t/\ell\) and let \(\ell\to\infty\).
With \(s=|\sin(\theta/2)|\) and
\(\gamma_\ell=2\arcsin(\alpha t s/(2\ell))\),
\(g_\ell^{\mathrm V}(-\cos\theta)=\sin(\ell\gamma_\ell)/(\ell\sin\gamma_\ell)\)
converges to \(\operatorname{sinc}(\alpha t s)=g_t^{\mathrm{cont}}(-\cos\theta)\),
with endpoint values taken by continuity. Thus the continuous filter
also gives positive semidefinite Gram matrices.

On the \(2K\) angles \(\theta_j=j\pi/K\), \(0\le j<2K\), the Gram
matrix with entries \(g(-\cos(\theta_i-\theta_j))\) is circulant.
Its eigenvalues divided by \(2K\) are nonnegative Fourier weights
summing to \(g(-1)=1\). Combining opposite frequencies and counting
the constant and Nyquist terms once gives cosine coefficients \(b_k\geq0\)
with \(\sum_k b_k=1\). These cosines interpolate the samples. Since
\(T_k(-\cos\theta)=(-1)^k\cos(k\theta)\), uniqueness of interpolation
gives \(a_k=(-1)^kb_k\). For the moments of~\eqref{eq:sm-moments}, the spectral
theorem and~\eqref{eq:sm-coeff-mass} therefore give
\[
 \left|\bm p^\dagger g(Y)\bm z-\sum_{k=0}^K a_k\widehat\mu_k\right|
 \leq \|g-I_Kg\|_{\infty,[-1,1]}
       +\max_{0\le k\le K}|\widehat\mu_k-\mu_k|.
\]
This bound applies to signed cross responses and to correlated moment errors.

\paragraph{Degree required over the full stiffness interval}
Set \(r=(1+y)/2\). For either filter at a lag \(t\leq T\), the
coefficient of \(r^m\) is bounded in magnitude by
\((\alpha T)^{2m}/(2m+1)!\). This follows from the sinc series in
continuous time. For matched Verlet, the expansion in Appendix~\ref{app:incoming-box}
gives the coefficient
\[
 \frac{(-\alpha^2(\Delta t)^2)^m}{\ell}\binom{\ell+m}{2m+1}
 =\frac{(-\alpha^2(\Delta t)^2)^m}{(2m+1)!}
    \prod_{i=1}^m(\ell^2-i^2),\qquad
 0\leq m\leq\ell-1,\quad t=\ell\Delta t .
\]
For \(y=\cos\theta\), the binomial expansion of
\(r^m=\cos^{2m}(\theta/2)\) has nonnegative cosine coefficients summing
to one. Lobatto interpolation aliases each higher Chebyshev mode to
one of degree at most \(K\), so \(\|I_K r^m\|_\infty\leq1\).
Since \(I_K\) reproduces the first \(K+1\) powers, the bias satisfies
\[
 \|g-I_Kg\|_\infty
 \leq 2\sum_{m>K}\frac{(\alpha T)^{2m}}{(2m+1)!}
 \leq 2\,4^{-(K+1)}e^{2\alpha T}.
\]
Thus \(K=O(1+\alpha T+\log(1/\eta))\) suffices for bias at most \(\eta/2\).

At fixed accuracy the linear dependence on \(\alpha T\) is necessary for
uniform approximation of the continuous lag at \(T\) over the full
stiffness interval. If \(b=\alpha T\geq\pi\), the filter \(g_T^{\mathrm{cont}}\) equals
one at \(y=-1\) and zero at \(y=-1+2\pi^2/b^2\). A degree-\(K\)
polynomial \(P\) of uniform error \(\eta<1/2\) therefore satisfies
\[
 \frac{(1-2\eta)b^2}{2\pi^2}\leq\|P'\|_\infty
 \leq K^2\|P\|_\infty\leq K^2(1+\eta)
\]
by Markov's inequality. A particular pair of probes can sample a
narrower spectral range.

\paragraph{Exterior initial data}
The incident terms in~\eqref{eq:shifted-both} use
\(\dot{\bm y}(0)\) and the load \(\bm k_0=A\bm y_0\)
from~\eqref{eq:incident-position}. For either vector \(\bm w\),
the additional moments \(C^\top T_k(Y)\bm w\) give estimates
\(\widehat{\bm b}_j(\bm w)\) of \(C^\top L_j(Y)\bm w\).
For a new exterior launch one supplies these incident vectors and
\(C^\top\bm y_0\). The material matrices \(\widehat G_j\) remain reusable.
Applying the node projection~\eqref{eq:sm-projection} to the incident
and driven terms gives the following extension of~\eqref{eq:sm-aux}:
\[
\begin{aligned}
 \bm s_{j,0}&=0,\\
 \bm s_{j,1}&=\Delta t\,\widehat{\bm b}_j(\dot{\bm y}(0))
       -\tfrac12 (\Delta t)^2[\widehat G_j\bm u_{\Gamma,0}
                          +\widehat{\bm b}_j(\bm k_0)],\\
 \bm s_{j,n+1}&=2\chi_j\bm s_{j,n}-\bm s_{j,n-1}
       -(\Delta t)^2[\widehat G_j\bm u_{\Gamma,n}
                          +\widehat{\bm b}_j(\bm k_0)],\qquad n\geq1,\\
 \widehat{\bm\tau}_n&=D_\times\bm u_{\Gamma,n}+C^\top\bm y_0
                          +\sum_{j=0}^K\bm s_{j,n}.
\end{aligned}
\]
For the graphene launch, \(\bm y_0=\dot{\bm y}(0)=0\), so
\(\bm s_{j,n}=-\widehat G_j\bm\xi_{j,n}\) in~\eqref{eq:sm-aux}.

\paragraph{Benchmark work counts}
The scattering comparison in Section~\ref{sec:incoming-test} assigns
an order-\(k\) moment the per-use work weight \(1+2k\).
For graphene (Section~\ref{sec:matched-work}), \(\alpha=19\);
\(M_0=C^\top C\) is computed directly, and orders \(1\) through \(14\)
are estimated. With 528 probe pairs, 63 repetitions, and phase length
\(2^{15}\), the setup uses
\[
 528\cdot63\cdot(2\cdot2^{15}-1)\sum_{k=1}^{14}2k
   =457{,}790{,}810{,}400
\]
ideal controlled walks. The reduced march in Table~\ref{tab:graphmatched}
uses one interface matrix--vector product per node per step, for 15 nodes.
\section*{Data and code availability}

The harmonic force constants and fitted graphene factor are available from Ref.~\cite{repository}.

\section*{Declaration of generative AI and AI-assisted technologies in the manuscript preparation process}
The author used OpenAI Codex for drafting, editing and literature searches. 

\setlength{\bibsep}{0.4pt plus .3pt minus .2pt}
\bibliographystyle{elsarticle-num}
\bibliography{refs}
\end{document}